\documentclass[aps,prx,twocolumn]{revtex4-2} 
\usepackage{graphicx}
\usepackage{amsmath,amsfonts,amssymb}
\usepackage{bm}
\usepackage{color}

\usepackage{xcolor}

\usepackage[normalem]{ulem}

\usepackage{soul}

\usepackage{graphics}

\begin{document}

\newcommand{\uu}{{\bf u}}
\newcommand{\etabf}{\mbox{\boldmath $\eta$}}
\newcommand{\he}{{\hat{{\bf e}}}}
\newcommand{\pp}{{\bf p}}
\newcommand{\Identity}{\boldsymbol{I}}
\newcommand{\F}{\boldsymbol{F}}
\newcommand{\rr}{{\bf r}}
\newcommand{\fvec}{{\bf f}}
\newcommand{\bq}{{\bf q}}
\newcommand{\vv}{{\bf v}}
\newcommand{\xx}{{\bf x}}
\newcommand{\NN}{{\bf \nabla}}
\newcommand{\hn}{{\hat{{\bf e}}}}
\newcommand{\hp}{{\hat{p}}}
\newcommand{\bfnabla}{\mbox{\boldmath $\nabla$}}
\newcommand{\bfz}{\mbox{\boldmath $\zeta$}}
\newcommand{\bnabrn}{{\bfnabla_{{\bf r}_n}}}
\newcommand{\bfxi}{\mbox{\boldmath $\xi$}}
\newcommand{\bnabr}{{\bfnabla_{{\bf r}}}}

\title{Hydrodynamics of simple active liquids: the emergence of velocity correlations}

\author{Umberto Marini Bettolo Marconi}
\affiliation{Universit\'a di Camerino, Dipartimento di Fisica, Via Madonna delle Carceri, I-62032 Camerino, Italy}
\author{Andrea Puglisi}
\affiliation{CNR-Istituto Sistemi Complessi, P.le A. Moro, I-00185, Rome, Italy}
\author{Lorenzo Caprini}
\email{lorenzo.caprini@gssi.it}
\affiliation{Universit\'a di Camerino, Dipartimento di Fisica, Via Madonna delle Carceri, I-62032 Camerino, Italy}

\date{\today}

\begin{abstract}
We derive the Hydrodynamics for a system of  $N$ active, spherical, underdamped particles, interacting through conservative forces.
At the microscopic level, we represent the evolution of the particles in terms of the Kramers equation for the probability density distribution of their positions, velocities, and orientations, while at a mesoscopic level we switch to a coarse-grained description introducing an appropriate set of hydrodynamic fields given by the lower-order moments of the distribution.
In addition to the usual density and polarization fields, the Hydrodynamics developed in this paper takes into account the velocity and kinetic temperature fields, which are crucial to understanding new aspects of the behavior of active liquids. 
By imposing a suitable closure of the hydrodynamic moment equations and  truncation of the Born-Bogolubov-Green-Kirkwood-Yvon hierarchy, we obtain a closed set of mesoscopic balance equations.
At this stage, we focus our interest on the small deviations of the hydrodynamic fields from their averages and apply the methods of the theory of linear hydrodynamic fluctuations.
Our treatment sheds light on the peculiar properties of isotropic active liquids and their emergent dynamical collective phenomena, such as the spontaneous alignment of the particle velocities. 
We predict the existence within the liquid phase of spatial equal-time Ornstein-Zernike-like velocity correlations both for the longitudinal and the transverse modes. At variance with active solids, in active liquids, the correlation length of the transverse velocity fluctuations is sensibly shorter than the length of the longitudinal fluctuations.
In particular, the latter depends on the sound speed and increases with the persistence time, while the former displays a weaker dependence on these parameters.
Finally, within the same framework, we derive the dynamical structure factors and the intermediate scattering functions and discuss how the velocity ordering persists in time. We find that the velocity decorrelates on a time-scale much longer than the one characteristic of  passive fluids.
\end{abstract}

\maketitle

\section{Introduction}
\label{intro}

In the last decade, significant progress has been made in the study of the collective  behavior of active (or also self-propelled) particles,  which comprise bacteria, cell assemblies, active colloidal suspensions, vibrated granular particles, autonomous micromotors, bird flocks, etc.~\cite{marchetti2013Hydrodynamics, elgeti2015physics, bechinger2016active}.
Understanding how to control and modify their unusual properties 
is of capital importance in many practical applications, and could revolutionize wide-ranging fields from medicine to robotics.
Since one of the possible practical applications of active particles could be self-assembly it would be important to
learn how to use them to obtain emergent materials and substances which have reliable, expected, and predictable properties.
From a thermodynamic viewpoint, active particles are systems out of equilibrium since they consume energy from the environment or internal chemical processes and generate mechanical persistent motion~\cite{gompper20202020}.
In statistical mechanics, this everlasting energy flow corresponds to a violation of the detailed balance condition.
Many  properties of Active matter are peculiar and absent in passive systems subject only to random thermal fluctuations.
Even in the absence of explicit attractive forces,
active particles may exhibit novel types of self-organization: they undergo a type of phase separation known as mobility induced phase separation (MIPS)~\cite{fily2012athermal, redner2013structure, buttinoni2013dynamical, cates2015motility, van2019interrupted}, crowd in the proximity of surfaces~\cite{solon2015pressure}, and form ``living crystals''~\cite{palacci2013living, mognetti2013living} that are mobile, break apart and reform again.

Besides the appearance of spontaneous density inhomogeneities, recent experimental, theoretical and numerical investigations have shown
the existence of spontaneous equal-time velocity correlations in  active matter systems, an emergent collective phenomenon.
This new property has been experimentally observed in different contexts, such as cell monolayers~\cite{garcia2015physics, henkes2020dense, sarkar2020minimal, alert2020physical} and bacterial colonies~\cite{dombrowski2004self, grossmann2014vortex, wioland2016ferromagnetic, peruani2012collective}. 
 Spontaneous velocity correlations represent a peculiar property of active matter systems and, thus, understanding the underlying physical mechanism could be an important advancement.
In most cases, these correlations have been explained by invoking velocity-aligning interactions which are the basic ingredient to produce the flocking transition in Vicsek-like models and Toner-Tu Hydrodynamics.
However, in some systems of self-propelled particles, it is possible to observe
fascinating velocity/activity patterns even in the absence
of this kind of interactions~\cite{grossmann2020particle}.
Henkes et al. and Caprini et al.~\cite{henkes2020dense,caprini2020spontaneous,caprini2020hidden} have shown that spatial velocity correlations do appear also in systems of spherical particles without alignment forces: this phenomenon is simply induced by the interplay between persistent active forces and steric repulsion.
The first research group compared experimental results observed in systems of high-density cell monolayers with a phenomenological theory~\cite{henkes2020dense}, while  the second group studied a suspension of two-dimensional active Brownian particles (ABP) under high-packing conditions \cite{caprini2020spontaneous} and put forward a microscopic theory predicting the exponential decay of the spatial velocity correlations. 
The characteristic coherence length was obtained in terms of the model parameters with~\cite{caprini2021spatial} or without  the effect of the inertia~\cite{caprini2020hidden}.
 Recently, Szamel and Flenner~\cite{szamel2021long} investigated active liquids and found numerical evidence of spontaneous velocity correlations. Such a study leads to the conclusion
that the emergence of self-organized patterns in the velocity field is a general property of active matter that may occur even in the absence of direct aligning interactions.

The existing theoretical treatments explain this phenomenon employing a microscopic approach where the evolution of the positions and velocities of each particle is explicitly considered. 
However, a systematic treatment based on coarse-grained collective variables is still lacking.
This paper aims to bridge the microscopic and  mesoscopic levels by developing a hydrodynamic theory of active liquids
able not only to reproduce the observed onset of longitudinal/transverse velocity correlations but also to predict new phenomena such as the slow relaxation of velocity fluctuations at large scales.

\begin{figure*}[t!]
\includegraphics[width=0.95\linewidth,keepaspectratio,angle=0]{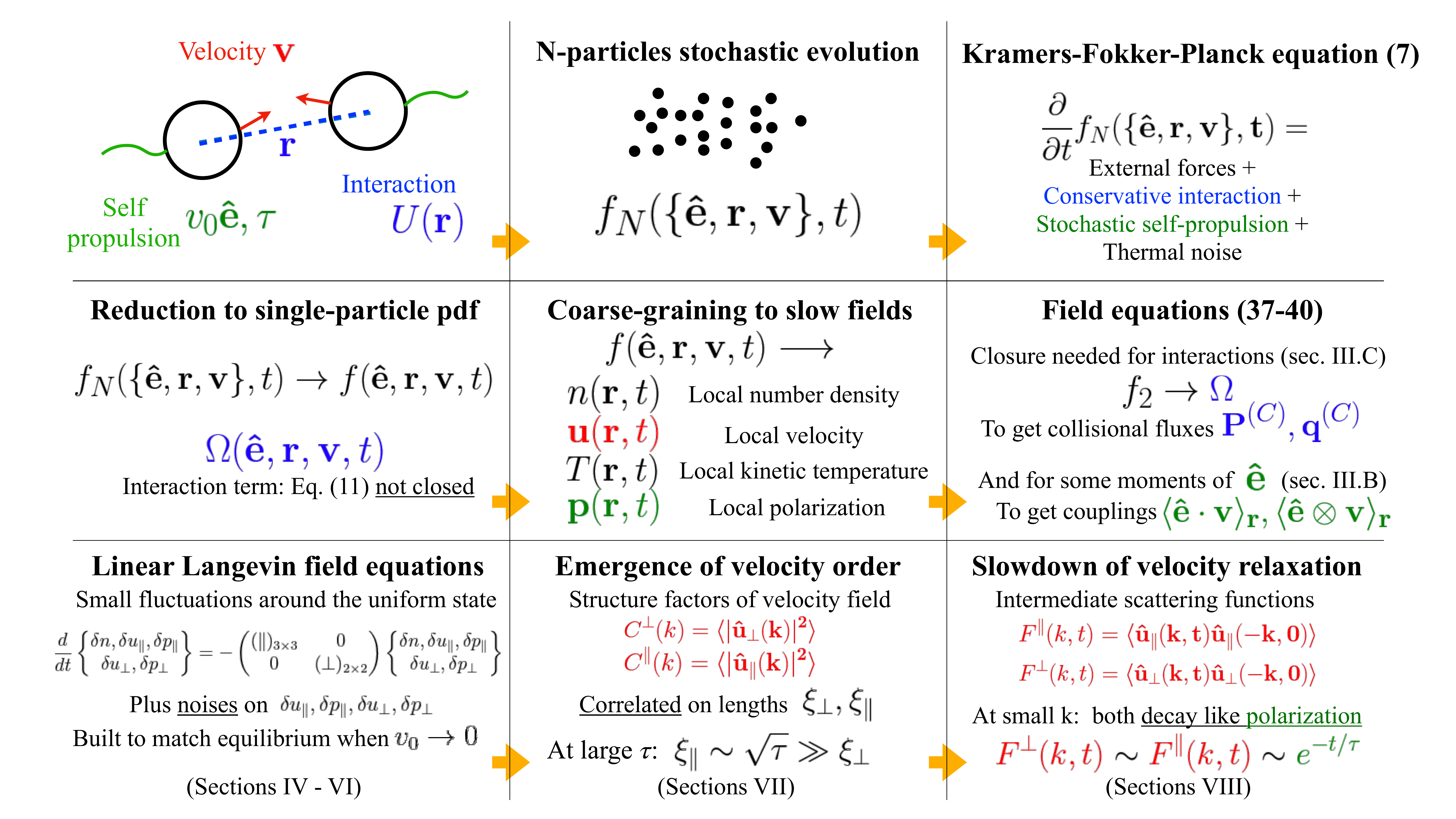}
\caption{A summary of the paper. The panels of the figure illustrate the main logical steps of this work, highlighting the most important observables considered and symbols introduced. 
They show the essential ingredients of the model (top frames), the hydrodynamic description in terms of coarse-grain fields, the closure approximations (middle frames), and the main results of the paper (last two frames in the third row). 
\label{fig0}}
\end{figure*}

The main actors and the logic line of our work are summarised in Fig.~\ref{fig0}.
Specifically, we consider a system of underdamped interacting active particles characterized by an active force following Active Brownian (ABP) or Active Ornstein-Uhlenbeck (AOUP) dynamics (for recent reviews, see Refs.~\cite{marchetti2016minimal, hecht2021introduction} and Ref.~\cite{martin2021statistical}, respectively).
These are minimal models for active matter, which combine mutual conservative interactions (e.g. repulsion for volume exclusion) and Brownian directed motion, but neglect hydrodynamic interactions or other kinds of explicit alignment forces.
 We first obtain a microscopic description in terms of the stochastic Kramers-Fokker-Planck (KFP)  evolution equation of the $N$-particle distribution function.  
Then, we connect the microscopic to the macroscopic by defining suitable hydrodynamic variables as averages (over the $N$-particle distribution) of local single-particle observables.
Finally, we derive from the KFP equation the relevant set of hydrodynamic balance equations which represent the first members of an infinite hierarchy involving the velocity and orientation moments of the distribution.

 While such a procedure is a standard tool in non-equilibrium statistical mechanics in the framework of passive systems~\cite{kreuzer1981nonequilibrium}, its implementation in the case of active matter offers new interesting challenges~\cite{julicher2018hydrodynamic}: due to the presence of the orientational degrees of freedom associated with the active force, the procedure not only entrains a larger number of fields, hence a different truncation of the hierarchy of equations for the moments, but also requires different approximations to treat the effective force which appears as a result of the interplay of active and repulsive forces.
The truncation of the hierarchy can be obtained by various procedures, either by heuristic methods or as
 in the case of the ABP model more systematically by the use of the Chapman-Enskog method as shown by Steffenoni et al.~\cite{steffenoni2017microscopic}. 
This leads to consider the dynamics of at least an extra field, the polarization, in addition to the 
usual fields of standard (passive) Hydrodynamics. 
Besides the truncation of the hierarchy, to obtain a closed set of equations, it is necessary to find a suitable representation of the interactions in terms of the hydrodynamic fields. 
In passive systems, it is well known,  as shown in the pioneering theoretical treatment of Irving and Kirkwood (IK)~\cite{irving1950statistical}, that (purely repulsive) interactions give rise to stress and energy flux contributions and viscous dissipation terms, in the equations for the momentum and energy densities, respectively. 
For a system of passive particles, the average internal force density can be expressed as the divergence of a stress tensor, providing a connection between statistical mechanics and continuum mechanics. 
In active systems, the peculiar interplay between repulsive forces and directed motion induces trapping of the particles, reducing their effective motility. 
Such a mechanism has been represented as an effective interaction that has no counterpart in passive fluids.
The understanding of this complex force, which has been the object of systematic investigation in the last two decades~\cite{cates2015motility,solon2015pressure,hermann2019phase,hermann2021phase}, is fundamental to obtain the correct hydrodynamic description of the active fluids.

We recall the existence of a rich literature where effective hydrodynamic equations are employed to understand collective phenomena in active systems. There are two main differences with respect to our treatment. 
First, most of these studies focus on models containing different ingredients, such as explicit alignment interactions (such as in Vicsek-like models) or anisotropies of the particles' shape. 
Second, the large majority of hydrodynamic studies concerning the ABP neglects the velocity field, typically involving only density and polarization: these fields account for the main macroscopic observed symmetry breaking phenomena, i.e. motility-induced-phase-separation and flocking transition. 
In many of such theories, the effect of the active force is represented by an effective ``active'' pressure term~\cite{speck2021coexistence} or an effective density-dependent local mobility~\cite{cates2015motility}. 

Neglecting the velocity fluctuations comes from the analogy with equilibrium models and the general argument that, in systems of passive particles dispersed in viscous fluids, such velocity fluctuations decay much faster than the density or polarization fluctuations, typically on a time-scale determined by the viscous drag.  
From the present study, we learn that this assumption should be reconsidered, as velocity and  polarization fluctuations relax on the same slow time-scale. 
However, from the theories, where the velocity fluctuations are neglected, we learn a few lessons which are crucial to close also our equations. For instance, in some steps of our derivation, we took inspiration from the recent work of Speck and co-workers~\cite{bialke2013microscopic, speck2020coexistence}: they considered a system of overdamped two-dimensional dry active disks (in the absence of hydrodynamic interactions) and by imposing the so-called force closure derived a self-consistent set of equations for the number and polarization densities~\cite{speck2015dynamical, speck2020collective}. Their approach captures the interplay between self-propulsion and repulsive interactions through an effective active force proportional to a density-dependent active speed whose form is similar to the one proposed by Cates and Tailleur in their seminal works~\cite{tailleur2008statistical, cates2013active} (see here for a review~\cite{cates2015motility}).

Our theory is suitable to describe isotropic active liquids and sheds light on their peculiar properties compared to those of usual passive liquids.
A fluctuating linear hydrodynamic approach around the homogenous state~\cite{kadanoff1963hydrodynamic,forster2018hydrodynamic} allows us to derive closed expressions for the spatial correlations of the velocity field. They originate from dynamical mechanisms that have no counterpart in equilibrium liquids, where the velocity field plays a marginal role only determining the relaxation towards the equilibrium.
Our theory is coherent with the scenario numerically observed in particle-based simulations both in solid and liquid configurations~\cite{caprini2020spontaneous, caprini2020hidden, henkes2020dense, szamel2021long}.
A particular outcome of the theory concerns the correlation lengths: we discover that the longitudinal velocity fluctuations are characterized by a correlation length which increases with the persistence time of the active force and is mainly determined by the sound speed. 
The transverse velocity fluctuations also display some degree of coherence, but the associated correlation length is much shorter, essentially because it depends on the shear, but not on the bulk modulus. 
We also investigate the spectrum of the hydrodynamic matrix as a function of the wavelength of the perturbation and, finally, we determine the dynamical structure factors and the intermediate scattering functions: from the latter we  
predict a pronounced slowdown of the velocity field, due to the activity.
To the best of our knowledge,
the latter point is a  central result that has not been highlighted in the  literature. At small values of the wavenumber, that is in a range of length-scales relevant for  macroscopic behavior, the fluctuation modes of the velocity field of a system of overdamped passive particles (such as colloidal suspensions) decay with a rate dictated by the viscous damping (normalized by mass). As demonstrated here, active liquids display a different behavior: the coupling with the polarization mode induces  an important reduction of the velocity field decay rate which is mainly determined by the inverse of the persistence time.  
At the end of this introduction, we anticipate a simplified discussion of such a general  slowdown mechanism, in order to bring to light its essential features.

The paper is organized as follows: in Sec.~\ref{Sec:model}, we introduce the stochastic model employed to describe a system of interacting active particles and develop the Born-Bogolubov-Green-Kirkwood-Yvon (BBGKY) hierarchy in the active case, while, in Sec.~\ref{Sec:hydro}, the hydrodynamic equations 
are derived, 
providing suitable closures for high-orders moments and the BBGKY hierarchy.
Secs.~\ref{eq:app_uniform} and~\ref{Sec:linearizedhydro} introduce a linearization procedure around the isotropic steady-state suitable to describe homogeneous active liquids, while Sec.\ref{Sec:fluctuatinghydro} employs the method of fluctuating Hydrodynamics to determine, in Sec.~\ref{equaltimecorrelations}, the spatial velocity correlation of transverse and longitudinal velocity modes.
 Finally, Sec.~\ref{dynamical} is dedicated to the study of the time-dependent properties of active liquids, showing results for the dynamical structure factor, the intermediate scattering function, and the eigenvalues of the hydrodynamic matrix as a function of the wavelength, both for transverse and longitudinal velocity modes.

\subsection{Summary of results: transverse and longitudinal velocity correlation and slowdown of the velocity field}

Many hydrodynamic theories for overdamped active particles neglect the velocity field since this is expected to undergo a fast decay likewise in systems of passive colloids in viscous solvents.
However, various arguments and analytical results from active particle models suggest that the self-propulsion induces an effective non-equilibrium memory term (effective inertia and space-dependent mobility) with a typical time determined by the persistence of the active force.
This description explains the slowdown of the dynamics due to the interactions but seems to be in contrast with the fast relaxation hypothesis of the velocity field assumed in some active hydrodynamic theories. 
To settle the question, in this paper, we derive coupled macroscopic equations for density $n(x,t)$, velocity ${\bf u}(\rr,t)$, polarization ${\bf p}(\rr,t)$ and kinetic temperature $T(\rr,t)$ fields (see Sec.~\ref{Sec:hydro} for definitions and balance equations), starting from a general microscopic active model with only conservative isotropic interactions.
In the second part of the paper, the theory is applied to homogeneous active liquids, resorting to linearized fluctuating Hydrodynamics.
As illustrated in Fig.~\ref{fig0}), a first conclusion concerns the existence of spontaneous order in the velocity field, already found in active solids, but here bearing new features: in particular, the correlation length of longitudinal modes, $\xi_\parallel$, is larger than that of transverse modes, $\xi_\perp$:
\begin{equation}
\xi_\parallel \approx v_s\sqrt{\frac{\tau}{\gamma}}  \gg \xi_\perp \,,
\end{equation}
and is mainly determined by the sound speed, $v_s$ (see Eqs.~\eqref{eq:xi_parallel_over} and~\eqref{eq:xi_perp_over} of Sec.~\ref{equaltimecorrelations}).
As a second important result, we show the slowdown of the velocity field: 
at a large spatial scale and in the overdamped regime, the time decay of  both velocity modes (longitudinal and transverse) is of the order of the persistence time, $\tau$, instead of being of the order of the inverse viscosity, $1/\gamma \ll \tau$ (see Sec.~\ref{dynamical}) as in passive suspensions. 
In the rest of this subsection, we discuss the essential mechanism underlying this slowdown effect.

Quite naturally, the active force  induces a coupling between the macroscopic velocity and polarization fields. This is clearly shown in Sec.~\ref{dynamical}, where dynamical correlations (the so-called intermediate scattering functions) are worked out in full generality both for transverse and longitudinal velocity modes. 
To give a flavor of the main underlying mechanism, we anticipate the case of zero-wavenumber ($k \to 0$) transverse modes.
In linear fluctuating Hydrodynamics, the transverse velocity and polarization modes at large scale, $V(t)=u_\perp (k=0,t)$ and $P(t)=p_\perp(k=0,t)$ respectively, obey the following Langevin equation:
\begin{equation}
\frac{d}{dt}\begin{pmatrix}
V\\P\end{pmatrix}=-\begin{pmatrix} \gamma & -\gamma \frac{v_0}{n_0}\\ 0 &\frac{1}{\tau}\end{pmatrix} \begin{pmatrix}V\\P\end{pmatrix} +  \begin{pmatrix}\sqrt{2 D_{V}}\eta_V\\\sqrt{2 D_{P}}\eta_P\end{pmatrix},
\end{equation}
where $v_0$ is the self-propulsion velocity, $n_0$ is the average number density of the active liquid, 
($D_{V}$, $D_{P}$) are effective diffusion coefficients (dependent on the model parameters), and ($\eta_V$, $\eta_P$) are independent white noises with  zero average and unitary variance. As illustrated in the rest of the paper, at finite $k>0$ the entries of the dynamical matrix and the noise amplitudes depend on $k$ through the transport coefficients of the theory. 
When $v_0=0$, a liquid of spherical passive colloids is recovered, the polarization becomes irrelevant and the velocity fluctuates around  $\langle V \rangle=0$ with its unique relaxation rate $\gamma$. 
When $v_0>0$, the eigenvalues of  the  dynamical matrix remain the same (as the determinant and the trace  of such a $2\times 2$ matrix are independent of $v_0$), therefore one would be tempted to conclude that $v_0$ does not affect the relaxation  of velocity fluctuations. Such a conclusion would be wrong as shown by the exact calculation of the correlation function:
\begin{equation}
\langle V(t) V(0) \rangle = A \exp(-\gamma t) + B \exp(-t/\tau),
\end{equation}
where $A$ and $B$ depend on the various parameters, see below, Eq.~\eqref{e104b}. 
In general, when $\gamma > 1/\tau$ (thus, in the overdamped regime) the coefficients satisfy $A \lesssim B$, therefore the first - rapidly decreasing - exponential becomes negligible in a time of order $1/\gamma$ and the decay is dominated by the slower relaxation of the second exponential. 

The mechanism summarised here gets more complicated in the case of the velocity longitudinal modes, which in addition to the polarization are coupled to the density fluctuations, but the result is similar: the dominant relaxation rate is $1/\tau$ also for these modes.
The non-trivial effect of non-equilibrium coupling between modes, even within a linear modelization, leading to slow relaxations and memory effects, has been studied in the past in the context of fluctuation-dissipation relations~\cite{PV09,VBPV09,CPV12,loos2020thermodynamic} and granular materials~\cite{sarra10b,gradenigo2011fluctuating,plati2020slow}.

\section{The model}\label{Sec:model}

We consider an assembly of identical active particles of  mass, $m$, mutually interacting through a short-range pairwise potential and immersed in a viscous solvent in a $d$-dimensional space.
The solvent  is quiescent and not affected by the particles' motion and is assimilated to a heat bath, acting as a source/sink of energy and momentum.
Each particle, identified by an index $n$, is subjected to an external time-independent force, $\fvec^{ex}_n=\fvec^{ex}(\rr_n)$, 
and a non-gradient force, $\fvec^{a}_n$, simply known as {\it active} force. 
 At such a level of description, we assume that $\fvec^{a}_n$ can be represented by a stochastic process with memory without specifying the detailed biological/physical mechanisms responsible for the active motion.
The active force is expressed as:
\begin{equation}
\frac{\fvec^{a}_n}{m}=\gamma v_0 \hn_n \nonumber\,,
\end{equation}where $v_0$ is the typical swim speed induced (in the absence of interactions or external forces) by the active force and $\gamma$ is the solvent viscosity. The term $\hn_n$ is a $d$-dimensional unit vector, representing the orientation of the active force. It evolves in time according to an unbiased random process with exponential memory having a typical correlation time, $\tau$.
The memory gives to the particle's trajectories the persistence character that is one of the salient features of active motion. 
In the framework of the continous stochastic processes, the most popular dynamics of $\hn_n$ are provided by
the Active Brownian Particles (ABP)~\cite{bialke2013microscopic, stenhammar2014phase, solon2015pressureNature, farage2015effective, caporusso2020motility, caprini2020hidden} and the active Ornstein-Uhlenbeck particle (AOUP) models~\cite{caprini2019activity, martin2020statistical, berthier2017active, wittmann2018effective, dabelow2019irreversibility, woillez2020active,flenner2020active}.
The latter has been often used in the theoretical studies (being simpler than the ABP) since it can reproduce the same phenomenology observed with ABP simulations, namely MIPS~\cite{fodor2016far, maggi2020universality} or accumulation near boundaries~\cite{caprini2018active, das2018confined}. 
Unlike Vicsek models or variants, it does not assume any explicit alignment interaction among the active forces of different particles.
In the ABP, the orientation is subject to the constraint $\hn^2=1$ and evolves according to the law: 
\begin{equation}
\frac{d\hn_n}{dt}= \sqrt{D_r}\bfz_n \times \hn_n \,,
\label{eq:ABPnoise}
\end{equation}
where $D_r$ is the rotational diffusion coefficient
and $ \bfz_n(t)$ is a white noise vector such that $\langle \zeta^\alpha_{n}(t) \zeta^\beta_{m}(s) \rangle  =  2\delta_{nm} \delta^{\alpha\beta} \delta(t-s)$. Here, latin and greek indices refer to particle numbers and spatial components, respectively.
The multiplicative noise in Eq.~\eqref{eq:ABPnoise} 
is treated according to the Stratonovich convention.
In the AOUP case, each $\hn_n$ evolves according to an Ornstein-Uhlenbeck process with correlation time, $\tau$:
\begin{equation}
\label{eq:AOUPactiveforce}
\frac{d \hn_n}{dt}  =-\frac{1}{\tau} \hn_n+ \sqrt{\frac{1 }{\tau}}  \bfz_n \,.
\end{equation}
In the dynamics~\eqref{eq:AOUPactiveforce} different Cartesian components of $\hn_n$ are uncorrelated and take on values ranging in the interval $[-\infty,\infty]$ and satisfy $\langle \hn^2 \rangle =d$.
Such a choice guarantees the correspondence with the ABP together with the relation $(d-1)D_r=1/\tau$ and a suitable rescaling of the orientation $\hn \to \hn/\sqrt{d}$~\cite{farage2015effective, caprini2019comparative}.
The equations of motion for underdamped active particles of mass, $m$, with position $\rr_n$ and velocity $\vv_n$, read~\cite{mandal2019motility, caprini2021spatial}:
\begin{subequations}
\label{kramers-b}
\begin{align}
&\frac{d\rr_n}{dt} = \vv_n  \\
 &\frac{d \vv_n}{dt}  =\frac{1}{m} \left[ 
\fvec^{ex}_n +\fvec^{a}_n
+ \mathbf{F}_n
\right]
-  \gamma \vv_n   + \bfxi_n \,,
\end{align}
\end{subequations} 
where $\bfxi_n$ is a white noise vector with unit variance.
Both the friction and $\bfxi_n$ originate from the solvent so that their amplitudes are related
by the fluctuation-dissipation relation:
\begin{equation*}
\langle \xi^\alpha_{n}(t)
\xi^\beta_{m}(s) \rangle  = 2 \gamma \frac{T_0}{m}\delta_{nm}
\delta^{\alpha\beta} \delta(t-s), 
\end{equation*}
where $T_0$ is the solvent temperature while, again, latin and greek indices refer to particle numbers and spatial components, respectively.
The force deterministic term, $\mathbf{F}_n$, is due to the interactions with the other particles and is expressed as
$$
\mathbf{F}_n=- \sum_{m(\neq n)} \bnabrn U(|\rr_n-\rr_m|) 
$$
being $U$ a generic pairwise repulsive potential accounting for volume exclusion effects.

\subsection{Many-particle Fokker-Planck description}

It is straightforward to derive 
the associated Kramers-Fokker-Planck~\cite{Risken}
evolution equation for the $9N$ dimensional phase-space probability 
density distribution $f_N=f_N(\{\hn,\rr,\vv\},t)$, where $\{ \rr,\vv\}$ indicates a $2\times d \times N$ dimensional phase space point
and $\{\hn_n\}$ the $d\times N$ space of the particle ``orientations":
\begin{eqnarray}&&
\left(\frac{\partial}{\partial t}+ \left[\vv_{n}\cdot\bnabrn+\left[\frac{\fvec^{ex}}{m}+
\frac{\fvec^{a}}{m}
\nonumber +\frac{1}{m} \mathbf{F}_n\right]\cdot \frac{\partial }{\partial \vv_n}  \right]
\right)f_N
\nonumber \\
&&\qquad=
 \gamma \frac{\partial }{\partial \vv_n} \left[  \frac{T_0}{m} \frac{\partial }{\partial \vv_n}+\vv_n  \right]f_N
+\frac{1}{ \tau} \mathcal{L}_af_N
\,.
\label{many4}
\end{eqnarray}
where repeated indices are summed.
Here, the operator $\mathcal{L}_a$ accounts for the dynamics of the active force and it is the only term containing the explicit dependence on the model considered (AOUP or ABP).
In particular, the operator $\mathcal{L}_a$ 
has the following form:
\begin{equation}
\label{eq:ActiveForce_evolutiongeneral}
\mathcal{L}_a f_N
=\frac{\partial }{\partial \hn_n} \cdot\Bigl(    \hn_n +  \frac{\partial }{\partial \hn_n}\cdot 
\boldsymbol{\mathcal{D}}_n\Bigl)f_N
 \,,
\end{equation}
where the difference between AOUP and ABP is contained in the form of the $d$-dimensional matrix $\boldsymbol{\mathcal{D}}_n$ for the $n$-th particle. Suppressing the particle index $n$, for convenience of notation, in the case of AOUP, we have $\boldsymbol{\mathcal{D}}_{AOUP}={\bf I}$, that is the identity matrix, while,
in the case of two-dimensional ABP, the matrix has the following spatial components~\cite{caprini2020active}:
\begin{equation}
\mathcal{D}_{ABP} =
\left( \begin{array}{cccc}
\hat{e}_y^2  & -\hat{e}_x\hat{e}_y  \\ 
-\hat{e}_y\hat{e}_x & \hat{e}_x^2 
\end{array} \right)\,.\nonumber\\
\label{D_ABP}
\end{equation}
The generalization to the three-dimensional case is straightforward and leads to similar results.
We remark that $\mathcal{L}_a$ is formed by two terms: i) a ``deterministic'' drag term (proportional to the first derivative with respect to $\hn$) common for ABP and AOUP and ii) a ``diffusive'' term (proportional to the second derivative with respect to $\hn$) that contains the only difference between ABP and AOUP. 
Moreover, we remark the following property that will be useful later: 
\begin{equation*}
\langle \boldsymbol{\mathcal{D}}_{ABP}\rangle \, d =  \boldsymbol{\mathcal{D}}_{AOUP} = {\bf I} \,,
\end{equation*}
 where the average is performed over all the variables of the system.
Therefore, to obtain the mapping between ABP and AOUP it is enough to rescale the ABP orientational vector $\hn$ with the constant factor $\sqrt{d}$, that is equivalent to map the ABP swim velocity onto $v_0\to \sqrt{d} \,v_0$. 
Bearing in mind this minor difference between AOUP and ABP, i.e. rescaling $\hat{\mathbf{n}}$ in the ABP case, the matrix $\boldsymbol{\mathcal{D}}$ satisfies the following tensorial relation:
\begin{equation}
\langle \boldsymbol{\mathcal{D}}\rangle =   {\bf I} \,.
\label{eq:matcalD_ABPAOUP}
\end{equation}
In the following, we discuss a theoretical approach that applies to both models since the differences between ABP and AOUP do not lead to major differences in the hydrodynamic description.
\subsection{Reduced description}

In order to proceed further as in the passive case, we define the reduced single-particle (marginal) distribution function, $f=f(\hn,\rr,\vv,t)$, as
\begin{equation}f(\rr,\hn,\vv,t)= \Pi_{n=2}^N\int d\hn_n d\vv_n d \rr_n
f_N(\{\rr,\hn,\vv\},t) \,.
\end{equation}By integrating out the $(\hn,\vv,\rr)$  coordinates of $(N-1)$ particles in Eq.~\eqref{many4}, we obtain the following non linear-equation
 \begin{eqnarray}&&
\Bigl(\frac{\partial}{\partial t}+\vv\cdot\bnabr+\frac{\fvec^{ex}}{m}\cdot  \frac{\partial }{\partial \vv}+
\gamma v_0\hn \cdot  \frac{\partial }{\partial \vv}  \Bigl) f
\nonumber \\&&\qquad=
\frac{1}{\tau}\mathcal{L}_a f
+\gamma  \frac{\partial }{\partial \vv} \Bigr( \frac{T_0}{m}  \frac{\partial }{\partial \vv} +\vv \Bigr) f
+ \Omega
\label{many4single}
\end{eqnarray}
where the interaction term, $\Omega=\Omega(\hn,\rr,\vv,t)$, is defined as:
\begin{equation}\Omega 
= \frac{1}{m} {\NN_{\bf v}} \cdot  \int   d\rr' \, d\vv' \, d\hn' \, 
f_2
{\bf \NN_{\bf r}}U(|\rr-\rr'|) \,,
\label{interaction0}
\end{equation}and involves the two-particle distribution function, $f_2=f_2(\rr,\hn,\vv; \rr',\hn', \vv',t)$, which is obtained from $f_N$
by integrating out $(N-2)$ particle coordinates.
 We remark that the relaxation of the system towards the steady-state occurs even in the absence of particle-particle couplings, due to the combined action of the solvent forces and self-propulsion. On the other hand, the interaction $\Omega$ not only contributes to the relaxation process via viscous  effects, heat transport, and polarization diffusion but is also responsible for the terms describing steric repulsion, effective attraction, and deviation of the local kinetic temperature from the heat bath temperature.

 Clearly, when the self-propulsion, $v_0$, vanishes, the distribution function factorizes
 into a translational and an orientational part
and Eq.~\eqref{many4} can be reduced to the Kramers equation describing
the inertial passive colloidal particles in equilibrium with a thermal bath. 

\section{Hydrodynamic balance equations}\label{Sec:hydro}
\label{Hydrodynamicsbalance}

To handle Eq.~\eqref{many4single}, which cannot be solved in general even when $\Omega=0$, we derive the evolution equations for a finite set of moments of the single-particle distribution function
(see also~\cite{steffenoni2017microscopic, klymko2017statistical}). 
The equations for the moments are obtained by multiplying Eq.~\eqref{many4single} by a  suitable number of products of the velocity, $\vv$ and of $\hn$ and integrating with respect to $\vv,\hn$ the evolution equation for the distribution. 
In contrast with the hydrodynamical treatment of passive fluids, where only $(2+d)$ variables are considered, namely, the number and momentum densities, and the kinetic temperature, the state space required to describe active systems is larger and it is necessary to include
the local polarization vector (that is the polarization of the active force). 

Even in the absence of interactions, the resulting set of balance equations forms part of an infinite hierarchy. As in the case of passive colloids, a truncation scheme is necessary.
Moreover, the presence of interactions in Eq.~\eqref{many4single}, brings in an additional difficulty  
because the $\Omega$-term contains, through $f_2$, the two-particle correlations. 
Thus, one is faced with two problems: a) break the moments' hierarchy and b)
obtain an expression of the interaction in terms of the moments
of the single-particle distribution and some known pair correlations. 

To develop an active hydrodynamic theoretical description, we introduce the following local fields (depending both on the position $\mathbf{r}$ and time $t$):
\begin{enumerate}
\item[i)]
The local  density, $n=n(\rr,t)$:
\begin{equation}n= \int d\vv\, d\hn \, f \,.
\label{density}
\end{equation}\item[ii)]
The local  velocity, $\uu=\uu(\rr,t)$, 
\begin{equation}n\uu=\int d\vv\, d\hn \,f\,\vv  \,.
\label{velocity}
\end{equation}\item[iii)]
The local kinetic temperature, $T=T(\rr,t)$,
\begin{equation} n T=
\frac{m}{ d}\int d\vv \, d\hn \,f\, (\vv-\uu)^2  \,.
\label{temperature}
\end{equation}where the Boltzmann constant is set equal to 1.
\item[iv)]
Following the current literature,
the  local polarization \cite{solon2015pressure} vector, $\mathbf{p}=\mathbf{p}(\rr,t)$ is: 
\begin{equation}\mathbf{p}=\int d\vv\, d\hn \, f\,\hn  \, .
\label{definizionepolarizzazione}
\end{equation}\end{enumerate}
The fields i), ii) and iii) are those usually accounted for by the Hydrodynamics of passive particles while the field iv) has an active origin.
In order to derive an adequate closure of the hydrodynamic equations, it is necessary to write the balance equation for at least two additional tensorial fields constructed combining the components of orientation and velocity vectors:
\begin{equation}\label{eq:tensor_ev_def}
n\langle\hn \otimes {\bf v} \rangle_r =\int d\vv\,  d\hn \, f \left[  \hn\otimes {\bf v} \right]
\end{equation}and
\begin{equation}\label{eq:tensor_ee_def}
n \langle\hn \otimes \hn \rangle_r
=\int d\vv\,  d\hn \, f \left[ \hn \otimes \hn \right] \,,
\end{equation}where the subscript $\langle \cdot \rangle_r$ indicates that the ensemble averages $\rr$ still depend on the position of the fluid element.
Here and in the following, the symbol $\otimes$ stands for the tensorial product, so that for instance $\hn \otimes \vv $ is the matrix of elements $\hat{e}_j \hat{v}_i$. Instead, the symbol $:$ will be used to denote the dyadic product between vectors or matrices.
We remark that the tensor $\langle\hn \otimes \hn \rangle_r$ is strictly related to the quadrupole tensor employed in the 
treatment of overdamped active particle systems ~\cite{winkler2015virial,speck2021coexistence}, whereas
$\langle\hn \otimes {\bf v} \rangle_r$ is a tensor whose trace is proportional to the work performed by the active force. 


\subsection{Balance equation for number, momentum, kinetic temperature densities}

 We derive a set of balance equations for number and momentum densities, and kinetic temperature  by projecting Eq.~\eqref{many4single} onto the Hilbert space spanned by the functions $1,\vv,m(\vv-\uu)^2/2$, respectively, as usual for passive colloids. 
This procedure immediately leads to the continuity equation by integrating Eq.~\eqref{many4single} over velocity and orientation degrees of freedom (in what follows, this procedure will be simply called ``integration''):
\begin{equation}\frac{\partial}{\partial t} n
+\nabla\cdot ( n \uu)=0 \, ,
\label{continuity}
\end{equation}that expresses the density conservation.
Here, we have used that $\int d\vv \, d\hn\,\Omega =0$, because the interaction conserves the number of particles.

Projecting onto the $\vv$-element, i.e. multiplying Eq.~\eqref{many4single} by $m\vv$ and integrating, leads to the momentum balance equation:
\begin{eqnarray}&&
m \frac{\partial}{\partial t} \left[n {\bf u} \right] +m\nabla\cdot \left[ n {\bf u} \otimes {\bf u}  \right] 
+\nabla \cdot {\bf P}^{(K)}- \nonumber\\
&&\qquad- m {\bf B}^{(v)}=n {\bf f}^{ex} +m\gamma v_0 {\bf p} -m\gamma n{\bf u} \,,
\label{hydromom}
\end{eqnarray}
while the equation for the kinetic temperature field is obtained by multiplying Eq.~\eqref{many4single} by $m(\vv-\uu)^2/2$. In that case, after not shown integrations and manipulations, usual for passive liquids, we obtain:
\begin{eqnarray}&&
\frac{d}{2} n \left(\frac{\partial}{\partial t} +{\bf u}\cdot \nabla \right) T+{\bf P}^{(K)} : \nabla {\bf u} +\nabla \cdot {\bf q}^{(K)}- B^{(vv)} \nonumber \\
&& = -\gamma d n (T-T_0) + \gamma  m v_0 \left( n \langle \hn \cdot  {\bf v}\rangle_r - {\bf p} \cdot {\bf u} \right)\, .
\label{hydrotemp}
\end{eqnarray}

Eqs.~\eqref{continuity},~\eqref{hydromom} and~\eqref{hydrotemp} are the minimal set of equations necessary to describe the Hydrodynamics of passive colloids and are expressed in terms of the so-called kinetic contributions and interaction terms.
The term ${\bf P}^{(K)}={\bf P}^{(K)}(\rr,t)$ represents the kinetic contribution to the pressure tensor and reads:
\begin{equation}{\bf P}^{(K)}=m\int d\vv \, d\hn\, f \left[({\bf v}-{\bf u})\otimes ({\bf v}- {\bf u})\right]  \,,
\label{kineticpressure}
\end{equation}while ${\bf q}^{(K)}={\bf q}^{(K)}(\rr,t)$ the one to the heat flux vector defined as:
\begin{equation}{\bf q}^{(K)}=\frac{m}{2}\int d\vv\, d\hn \, f \left[(\vv-\uu)^2(\vv-\uu)\right] \,.
\label{qk}
\end{equation}
Finally, Eqs.~\eqref{hydromom} and~\eqref{hydrotemp} contain the interactions terms (simply called $B^{(\cdot)}$) that are those involving the two-body distribution through $\Omega$. These terms account for the interactions among particles and are defined as:
\begin{equation}
{\bf B}^{(v)} =\int d\vv \, d\hn\,\Omega \,{\bf v}  \,,
\end{equation}
and
\begin{equation}
 B^{ (vv)} = \int d\vv \, d\hn\, \Omega\,\frac{m}{2}(\vv-\uu)^2 \,.
\end{equation}
In Eq.~\eqref{hydromom} the second, third, and fourth terms in the l.h.s. represent a momentum flow. 
In the r.h.s., $n\,\mathbf{f}^{ex}$ and $\gamma v_0 {\bf p}$ are two sources of momentum due to the presence of the external field and the active force, respectively. 
Finally, $-m\gamma n {\bf u}$ represents the  average solvent drag force density which opposes the motion in the direction of the velocity,
while term  $-m {\bf B}^{(v)} $ is the {\it internal force density}. 
It represents the rate of change of the momentum of the active particles induced by their mutual interactions.
In passive systems, the Irving-Kirkwood theory \cite{irving1950statistical} identifies such a force with the divergence of the pressure tensor, ${\bf P}$ (the negative of the stress tensor) according to
\begin{equation}\nabla \cdot  {\bf P}^{(C)}=-m {\bf B}^{(v)}  \,.
\label{IrvingK}
\end{equation}${\bf B}^{(v)}$, which vanishes in the bulk, i.e. under homogeneous conditions, is a well-studied quantity in the case of passive particles.
In the case of spherically repulsive interparticle interaction potential and
when ${\bf u}=0$,
the term ${\bf B}^{(v)}$ describes the repulsion that is solely due to the inhomogeneous density distribution.
When the fluid velocity is inhomogeneous, in addition to 
the hydrostatic  contribution to the pressure, ${\bf B}^{(v)}$ contains also a
viscous contribution~\cite{marconi2009kinetic,marconi2010dynamic}. 
The Irving-Kirkwood procedure~\cite{irving1950statistical} gives an explicit expression to derive this term.
When all terms on the right hand side of Eq.~\eqref{hydromom} are set to zero, i.e. when the friction, the active force, and the external force are suppressed, such an equation contains the same physics as the Navier-Stokes equation (NSE) describing a  compressible viscous fluid in motion. In fact, in the limit of small gradients, the pressure contributions can be expressed in terms of the static pressure and the dynamical stress tensor via the strain rates.
The dissipation contained in the NSE is only due to the internal friction of the fluid, while here two additional mechanisms of injection and dissipation of energy are present. 
The first one, typical of colloidal suspensions, corresponds to the interaction with the solvent, causing momentum suppression and energy reinjection.
The second mechanism is the active force which injects momentum through the term $\gamma v_0 {\bf p}$ but randomizes orientation since $\hn$ follows a stochastic evolution (ABP or AOUP models). 

The kinetic temperature field $T$ describes how the variance of the velocity distribution depends on the interactions and the active force. The second, third and fourth term in the l.h.s. of Eq.~\eqref{hydrotemp} are analogous to those  of a passive fluid and 
represent compressional work and heat transport. 

 In particular, ${\bf P}^{(K)}$ and ${\bf q}^{(K)}$ are the kinetic contributions to the pressure tensor and to the heat flux, respectively. 
Similarly, the $B^{(vv)}$ terms, representing the rate of change of kinetic energy induced by the interactions, can separated into a collisional heat flux contribution and a term accounting for the viscous and compressional work~\cite{kreuzer1981nonequilibrium,lautrup2011physics}:
\begin{eqnarray}B^{(vv)} &&= 
\int d\vv \, d\hn\,  \frac{(\vv-\uu)^2}{2} \NN_{\bf v} \cdot \int d\rr'\, d\vv' \, d\hn'
f_2 {\bf \NN_{\bf r}}U \nonumber \\
&&=- {\bf P}^{(C)}:\nabla\uu-\nabla \cdot {\bf q}^{(C)}  \,.
\label{bc2}
\end{eqnarray}
 Such an equation defines the divergence of the field ${\bf q}^{(C)}$, the "collisional component" of the heat flux, which in the high-density regime becomes dominant over ${\bf q}^{(K)}$ .
 In the energy equation Eq.~\eqref{hydrotemp}, the effect of the active force is encapsulated in the last parenthesis.
This term is proportional to $v_0$ and has an energetic interpretation: it is nothing but the average work density (performed by the active force) and acts as an additional source of energy, usually larger than $T_0$, i.e. the contribution of the solvent. 

\subsection{Balance equation for the polarization field and breaking of the moment hierachy}

 At variance with the Hydrodynamics of passive particles, the momentum balance equation~\eqref{hydromom}
and the temperature equation~\eqref{hydrotemp} involve two new fields, the polarization, $\mathbf{p}$ and   $\langle {\hn} \cdot {\bf v}
\rangle_r$, respectively.
 To fix them we enlarge the set of hydrodynamic equations  by projecting 
the distribution $f$ over $\hn$ and $\hn \otimes \vv$ and obtain two additional relations.
First,
multiplying Eq.~\eqref{many4single} by $\hn$ and integrating, we obtain  balance equation for the polarization:
\begin{equation}\frac{\partial}{\partial t} {\bf p}+\nabla \cdot  \left(n\langle {\bf v} \otimes \hn\rangle_r \right)=-\frac{1}{\tau}  {\bf p} \, ,
\label{hydropol}
\end{equation}where we have used that $\int d\vv d\hn\,\Omega\,  \hn=0$.
Equation~\eqref{hydropol} explicitly depends on the two-body cross-correlation between particle velocity and active force $v_0\langle {\bf v}\otimes\hn\rangle_r$ and does not depend on the choice of the active force, being the same both for ABP and AOUP models.
As shown by Cates \& Tailleur, the  equation for the polarization brings in the key ingredient responsible for MIPS. 
Equation~\eqref{hydropol} reveals that the local polarization changes for two reasons: the advection associated with the incoming flux of particles carrying the active force with them,  and a sink term due to the reorientation of the polarization after a typical time $\tau$. 
The form of Eq.~\eqref{hydropol} suggests that on a time scale $t<\tau$, $ {\bf p}$ remains approximately constant and 
 the polarization can be expressed as the divergence of the tensor $n \langle {\bf v} \otimes \hn\rangle_r $.
To determine the latter quantity, we need to consider the polarization flux equation,
which is obtained by multiplying Eq.~\eqref{many4single} by $ {\bf v}\otimes \hn$ and integrating: 
\begin{eqnarray}&&
\frac{\partial}{\partial t} [n \langle {\bf v} \otimes \hn\rangle_r ]+\frac{\partial}{\partial r_k} \left( n \langle v_k {\bf v} \otimes \hn \rangle_r\right)
-{\bf B}^{({\hat e} v)}  
= \frac{{\bf f}^{ex}}{m} \otimes {\bf p} \nonumber\\
&&+ v_0 \gamma n \langle \hn\otimes \hn\rangle_r+\frac{1}{\tau} {\bf u} \otimes {\bf p} 
-(\gamma+\frac{1}{\tau}) n \langle {\bf v}\otimes \hn\rangle_r \,,
\label{tensorbalance}
\end{eqnarray}
where the interaction contribution ${\bf B}^{({\hat e} v)}={\bf B}^{({\hat e} v)} (\rr,t)$ is defined by the tensor 
\begin{equation}
{\bf B}^{({\hat e} v)}  = \int d\vv d\hn\,\Omega \, \hn \otimes{\bf v} \,.
\label{def:Cnv}
\end{equation}
This term represents the highly non-trivial contribution of the interaction operator $\Omega$ to the ${\hat e}_i  v_j$-moment equation.
It describes the combined effect of self-propulsion and steric repulsion and, in recent papers, it has been termed ``indirect interaction''~\cite{solon2015pressure}.
Again, Eq.~\eqref{tensorbalance} does not depend on the ABP or AOUP choice. 

To close the moments hierarchy, we consider the equation for, $\langle \hn \otimes \hn \rangle_r$, simply by multiplying Eq.~\eqref{many4single} by $\hn \otimes \hn$ and integrating:
\begin{flalign}
\frac{\partial}{\partial t} & [n ( \langle \hat \hn \otimes\hn \rangle_r -{\bf I})]+\nabla \cdot  \left[n \langle \vv \otimes \hn \otimes \hn\rangle_r  \right] - {\bf I} \,\nabla \cdot \left(n\uu \right) \nonumber\\
&=-\frac{2}{\tau}   n ( \langle \hn \otimes \hn\rangle_r -\langle \boldsymbol{ \mathcal{ D}}\rangle_r)
\label{nnmom1}
\end{flalign}
By factorizing the averages of the products of three operators in Eq.~\eqref{nnmom1}, neglecting the third order cumulant and using the continuity equation,
we find the following relation between steady averages:
\begin{flalign}
  \label{nnmom3}
n \langle \hn \otimes \hn \rangle_r=& n \langle\boldsymbol{\mathcal{D}}\rangle_r-
\frac{\tau}{2} \nabla \cdot  \left( n \uu \otimes\langle  \hn \otimes \hn\rangle_r \right)\\
&-\frac{\tau}{2} \nabla \cdot \left[ \langle \vv \otimes \hn \rangle_r \otimes \mathbf{p}  +  \langle \vv \otimes \mathbf{p} \otimes \hn \rangle_r \right]\nonumber \,.
  \end{flalign}
By neglecting the gradient terms in Eq.~\eqref{nnmom3}, we get  the simpler approximation for $\langle \hn \otimes \hn \rangle_r$:
\begin{equation}
\langle \hn \otimes \hn\rangle_r = \langle\boldsymbol{\mathcal{D}}\rangle_r \approx {\bf I}\,.
\label{eq:ninj_approx}
\end{equation} 
 In the AOUP model, Eq.~\eqref{eq:ninj_approx} is exact, while in the case of ABP, we have used a further approximation valid in the homogeneous phases, $\langle\boldsymbol{\mathcal{D}}\rangle_r \approx \langle\boldsymbol{\mathcal{D}}\rangle$. Now, Eq.~\eqref{eq:matcalD_ABPAOUP} (which holds also for ABP, after a suitable variable rescaling) leads to the result.
Further expressions involving gradient terms in the steady-state of  $\langle \hn \otimes \hn \rangle_r$ have been derived for systems of overdamped ABP~\cite{bertin2006boltzmann}. 
These more refined treatments of $\langle \hn\otimes\hn\rangle_r$ could be used in our theory, but, here, we have 
employed the simplest closure since we will consider only homogeneous rather than phase-separated configurations.

To estimate the third-order moment in Eq.~\eqref{tensorbalance}, we employ a Gaussian ansatz analogous to the one employed to obtain Eq.~\eqref{nnmom3}: 
\begin{eqnarray}
&&
n \langle \hat \vv\otimes \vv\otimes\hn  \rangle_r \approx   {\bf I}\otimes {\bf p}\frac{T}{m} -    \uu\otimes\uu\otimes{\bf p} \\
&&+n \langle \vv\otimes\uu\otimes\hn \rangle_r+ n \uu\otimes\langle \vv\otimes\hn \rangle_r \nonumber
\end{eqnarray}
so that the following relation holds:
\begin{eqnarray}
&&\nabla\cdot \left( n \langle \vv \otimes \vv\otimes \hn\rangle_r\right)
\approx \nabla \otimes {\bf p}\frac{T}{m}  -  \nabla \cdot \uu \otimes {\bf u}\otimes {\bf p}   \nonumber\\
&&+ \nabla \cdot  \left[ n \langle \vv \otimes \uu \otimes\hn \rangle_r + n \uu \otimes \langle \vv \otimes \hn  \rangle_r\right] \,.
\label{eq:thirdordermoment}
\end{eqnarray}
Plugging the approximation~\eqref{eq:ninj_approx} in the steady-state version of Eq.~\eqref{tensorbalance} and using Eq.~\eqref{eq:thirdordermoment}, we get the following relation for $\langle \hn \otimes \vv\rangle_r$: 
\begin{eqnarray}
&&
 n \langle \vv \otimes \hn\rangle_r =\frac{\tau}{1+\tau\gamma}  \Biggl[  \gamma n v_0 {\bf I} + {\bf B}^{({\hat  e} v)}
+ \left({\bf f}^{ex} -\nabla T\right)\otimes \frac{{\bf p}}{m} \\
&&+ \uu \otimes \frac{{\bf p}}{\tau}  - \nabla \cdot \bigl(n \langle\hat \vv \otimes \uu \otimes \hn \rangle_r + n \uu\otimes \left[\langle\hat \vv\otimes \hn  \rangle_r  - \uu\otimes {\bf  p}	\right]  \bigr)
\Biggr] \nonumber \,.
\label{rlm2_2}
\end{eqnarray}
which is our closure of the moment-hierarchy. 
\begin{widetext}
Summarizing the results of this section, we have obtained a set of balance equations 
for the first moments of the distribution function:
\begin{subequations}
\begin{align}
&
\partial_{t}n = - \nabla\cdot(n {\bf u}) \label{v61}\\
&
\partial_t T + {\bf u}\cdot\nabla T+\frac{2 m}{n d}\gamma v_0  \uu \cdot{\bf p}
= -\frac{2 }{nd}(\nabla\cdot {\bf q}+ {\bf P}:\nabla{\bf u}) 
+ 2\gamma(T_0 - T)+ \frac{2\gamma}{d}
m v_0   \langle \hn \cdot\vv\rangle_r \label{eq:T_together}\\
&
\partial_{t}{\bf u} + {\bf u}\cdot\nabla{\bf u} = - \frac{1}{m n}\nabla\cdot {\bf P} - \gamma{\bf u}+ \frac{{\bf f}^{ex}}{m} +\frac{1}{n}\gamma v_0 {\bf p}\label{he.1}\\ 
&
\partial_{t} {\bf p} + \nabla\cdot  (n \langle \hn \otimes \vv \rangle_r)=-\frac{1}{\tau} {\bf p} \,.
\label{lasteq}
\end{align}
\end{subequations}
where the symbols ${\bf P}$ and ${\bf q}$ are a compact notation for ${\bf P}^{(K)}+{\bf P}^{(C)}$ and ${\bf q}^{(K)}+{\bf q}^{(C)}$, respectively, and the term $\langle \hn\otimes\mathbf{v} \rangle$ and $\langle \hn\cdot\mathbf{v} \rangle$ can be obtained by Eq.~\eqref{rlm2_2}.
\end{widetext}
To proceed further, we need to develop a suitable closure procedure to express the pressure, heat flux and $\mathbf{B}^{(\hat e, v)}$ as a function of the hydrodynamic moments and their gradients. 
This task will require the estimate of the various collisional contributions.

\subsection{Closure of the BBGKY hierarchy for active particles}

Since $\Omega$, given by Eq.~\eqref{interaction0}, contains the two-particle distribution function,
$f_2$, it is necessary to have an approximation to determine how it can be expressed in terms of the one-particle and two-particle properties of the system such as the configurational correlation function.
For passive systems, there are several approximations concerning $f_2$ ranging from the Boltzmann {\it Stosszahl-ansatz}~\cite{cercignani1988boltzmann} to the Enskog-like theories~\cite{bellomo1991mathematical}, to the Kirkwood superposition approximation~\cite{hansen1990theory}.
They are not exact but, in many cases of interest,
lead to fairly good predictions both for hydrodynamic properties. 
In active systems, the dependence of the distribution functions on the orientation vector $\hn$ poses an extra challenge to the effort of finding an accurate approximation for $f_2$.
However, it is possible to resort to phenomenological arguments to determine the form of ${\bf B}^{({\hat e} v)}$.
The seminal idea was conceived by Tailleur and Cates~\cite{cates2015motility} and elaborated by Speck~\cite{speck2020collective} and more recently  by de Pirey et al.~\cite{de2019active} within a model of active hard-spheres in infinitely many dimensions. 
According to Ref.~\cite{speck2020collective}, in a homogeneous system of ABP, the conditional probability of finding
a second particle a distance $r$ from a first particle fixed at the origin and having a fixed orientation $\hn$ is axisymmetric with respect to $\hn$
and gives rise to a force that opposes direct motion. 
Such an effect due to the interplay between the self-propulsion and the repulsive force can be represented as a force parallel to $\hn$ but having a renormalized propagation speed: its value changes from $v_0$, typical of isolated active particle, to a density-dependent value~\cite{stenhammar2013continuum}: 
\begin{equation}
\label{eq:v(n)_phenomenological}
v[n]=v_0 \left(1-\frac{n}{n_c}\right) \,,
\end{equation}
where $n_c$ is a constant with the dimension of a density that is a function of the model parameters.
How does this phenomenological theory translate into the formalism of this paper? Is it possible to choose a particular form of $f_2$ 
which is consistent with Eq.~\eqref{eq:v(n)_phenomenological}?
We give the following argument: let us consider first the single-particle distribution $f$ and then proceed to guess the form of the $f_2$ distribution function.
Supposing that the first four hydrodynamic moments are known, we can construct a trial single-particle distribution
having these moments by the following ansatz:
\begin{eqnarray}f_{trial}=\phi_M  \psi_0\bigl[n+ {\bf p} \cdot \hn + \dots\bigr]
\label{trialddf}
\end{eqnarray}
where $\phi_{M}=\phi_M(\vv,\rr,t)$ is the local Maxwellian distribution that is defined by
\begin{equation}
\phi_M=\frac{m^{d/2}}{(2\pi T)^{d/2}}\exp\left(-\frac{m}{ T} \frac{(\vv-\uu)^2}{2}\right)
\end{equation}and $\psi_0=\psi_0(\hn)$ reads:
\begin{equation}\psi_0=\frac{1}{(2\pi )^{d/2}}e^{-\frac{1}{2 } \hn^2} \,.
\label{eq:psi0_hatn}
\end{equation}We remark that, in Eq.~\eqref{trialddf}, we have neglected deviations from the local Maxwellian velocity distribution apart from those stemming from the polarization. This assumption can be questioned in the case of phase separation, while it is reasonable in homogeneous configurations.
However, still in the latter case, the largest effect of the $\hn$ field is to distort the distribution function from its Maxwellian form.

Now, we turn to the two-particle distribution and, taking inspiration from Eq.~\eqref{trialddf}, we approximate $f_2=f_2(\rr,\hn,\vv;\rr',\hn',\vv',t)$ by the following form:
\begin{eqnarray}f_2=&&\phi_M  \psi_0  \phi'_M \psi'_0 \times
 \nonumber\\&&
 \times\Biggl[n n' g_2
+ \hn\cdot \langle {\bf m}\rho'\rangle
+ \hn' \cdot \langle{\bf m}'\rho \rangle+
\dots \Biggr]  \,,
\label{f2approx}
\end{eqnarray}
where the prime means that the observable is calculated at a phase-space point $(\mathbf{r}', \mathbf{v}', \hn')$  that differs from $(\mathbf{r}, \mathbf{v}, \hn)$.
The fields $\rho=\rho(\rr,t)$ and ${\bf m}={\bf m}(\rr,t)$ are defined as:
\begin{flalign}
&\rho=\sum_n^N \delta(\rr-\rr_n(t)) \,,\\
&{\bf m}\equiv \sum_n^N \hn_n \delta(\rr-\rr_n(t))
\end{flalign}
while $g_2=g_2(\rr,\rr',t)$ represents the configurational pair correlation function.

Once the form of $g_2(|\rr-\rr'|)$ is fixed, using the approximation~\eqref{f2approx} and the specific form of $\Omega$ given by Eq.~\eqref{interaction0}, it is possible to obtain closed expressions for the two collisional terms $B_i^{(v)}$ and $B^{(vv)}$. 
Such a task, normally performed in the study of passive fluids with repulsive interactions, is the subject of a vast literature~\cite{van1973modified,karkheck1981kinetic,bellomo1991mathematical,santos1998kinetic}. 
In the following, we shall approximate $B_i^{(v)}$ and $B^{(vv)}$ by the corresponding quantities of an elastic hard-disk system. 
Instead, the evaluation of the $ B^{({\hat e} v)}$ integral requires the knowledge of orientational correlations which are not normally considered in the studies of simple passive liquids and involves the correlation term proportional to $ \hn\cdot \langle {\bf m}(\rr,t)\rho(\rr',t)\rangle$ in the square parenthesis of Eq.~\eqref{f2approx}. 
After defining the following marginal correlation:
\begin{eqnarray}&&
n n' G_2\equiv \int d\vv \, d\vv' f_2\approx  \nonumber\\&&\psi_0  \psi'_0 \times
\Biggl[n n' g_2
+ \hn\cdot \langle {\bf m}\rho'\rangle
+ \hn' \cdot \langle{\bf m}'\rho\rangle+
\dots \Biggr]\nonumber \,
\end{eqnarray}
it is possible to rewrite Eq.~\eqref{def:Cnv} as
\begin{equation}m \mathbf{B}^{({\hat e} v)}
=- n \int d\hn \int d\hn'  \, d\rr'  n' G_2\, \left[\hn\otimes \nabla U\right]  \,,
\label{Cnv0b}
\end{equation}where we have integrated over the velocity degrees of freedom.
To go further, we need a prescription to evaluate the integral in Eq.~\eqref{Cnv0b} containing the pair correlation function $G_2$, whose detailed form is unknown. 
As indicated by our parametrization, $G_2$ contains a spherically symmetric part $g_2$  plus a contribution that depends
on the angles associated with $\hn$, $\hn'$ and the vector $\rr'-\rr$. 
By symmetry arguments one sees that only the $ \hn\cdot \langle {\bf m}(\rr,t)\rho(\rr',t)\rangle$ term contributes to the integral in Eq.~\eqref{Cnv0b}.
To obtain an expression for $B^{({\hat e} v)}$, we approximate the integral in Eq.~\eqref{Cnv0b} with the help of a semi-empirical formula that has been proved for active hard-spheres in infinitely many dimensions~\cite{de2019active}:
\begin{eqnarray}&&
\int d\hn
\int d\hn'  d\rr'  
n' G_2\, \left[\hn\otimes \nabla U\right] \approx m\gamma v_0\frac{n}{n_c} \, {\bf I} \,.
\label{eq:DePirey_approx}
\end{eqnarray}
Such an equation represents the integral of the interparticle force with the correlation of a particle with polarization $\hn$ at $\rr$  with a second particle at $\rr'$~\cite{speck2020collective,de2019active}.
The formula can be understood as follows: by balancing the active force with the repulsive force we obtain the typical scaling of the potential
$ \nabla U(|\rr-\rr'|) \approx m\gamma v_0 \hat{\mathbf{e}}$. 
On the other hand, the pair correlation $G_2$ is very peaked when  the two following conditions are satisfied: a) $|\rr-\rr'|/\sigma\sim 1$ (where the two particles are at the distance of closest approach being $ \sigma$ their effective diameter) and b) $(\rr'-\rr)\cdot(\hn'-\hn)< 0$ corresponding to the case of a pair of colliding particles.
On the contrary, $G_2$ is depleted when two particles move apart.

To conclude, by inserting Eq.~\eqref{eq:DePirey_approx} into Eq.~\eqref{Cnv0b}, we obtain an approximate but explicit representation of the collisional term:
\begin{equation}{\bf B}^{({\hat e} v)}\approx -\gamma v_0\frac{n^2}{n_c} {\bf I} \,,
\label{pinteraction}
\end{equation}
This result is consistent with the {\it force closure} employed and recently reviewed by Speck~\cite{speck2020collective}, in the simpler case of homogeneous configurations.
Notice, that for a passive system, $v_0=0$, this term vanishes being proportional to the active force strength $\gamma m v_0$. 
Moreover, the approximated expression for ${\bf B}^{({\hat e} v)}$ depends quadratically on the local density and is negative.
Remarkably, $\mathbf{B}^{(\hat e v)}$ is a non-gradient term, 
resulting from the correlations between particles orientation and the pairwise repulsive force.

Replacing $\mathbf{B}^{(\hat e v)}$ given by Eq.~\eqref{pinteraction} in Eq.~\eqref{rlm2_2} , we obtain the following self-consistent equation for the elements of the tensor $\langle\hn\otimes\vv\rangle_r$:
\begin{eqnarray}
\label{rlm2_2bis}
&&
 n \langle \vv \otimes \hn\rangle_r =\frac{\tau}{1+\tau\gamma}  \Biggl[  \gamma n v[n] {\bf I}
+ \frac{{\bf f}^{ex}}{m}\otimes {\bf p} -\nabla \otimes {\bf p}\frac{T}{m} \\
&&+ \uu \otimes \frac{{\bf p}}{\tau}  - \nabla \cdot \bigl(n \langle\hat \vv \otimes \uu \otimes \hn \rangle_r + n \uu\otimes \left[\langle\hat \vv\otimes \hn  \rangle_r  - \uu\otimes {\bf  p}	\right]  \bigr)
\Biggr] \nonumber \,.
\end{eqnarray}
Such a result has a simple physical interpretation~\cite{stenhammar2013continuum}:
at finite densities, collisions slow particles down,  and reduce the propulsion speed with respect to the value, $v_0$,
of an isolated particle.
At each collision, the combined effects of persistent motion and excluded volume 
lead to a temporary immobilization of a particle lasting a time $\tau_C\leq \tau$. Thus, $N_C$ collisions occurring during the propagation
time $\tau$ at constant speed $v_0$ reduce the effective distance (i.e. the persistence length) travelled during the persistence time $\tau$, from $\ell=v_0 \tau$
 to $\ell'=v_0 (\tau-N_C \tau_C)$.
 The number of collisions, $N_C$, is roughly given by $\ell/\ell_{MF}$ , where the mean free path $\ell_{MF}\sim1/n\sigma$. 
Therefore, the effective density-dependent propulsion speed becomes $v[n]=\ell'/\tau=v_0 (1-v_0 \sigma \tau_C n)$ in agreement with Eq.~\eqref{eq:v(n)_phenomenological} upon fixing $n_c\approx v_0 \sigma \tau_C$. 

\subsubsection{Estimate of the collisional terms ${\bf B}^{(v)}$ and $B^{(vv)}$}
 We approximate the terms ${\bf B}^{(v)}$ and $B^{(vv)}$, and the related tensors $\mathbf{q}$ and $\mathbf{P}$ featuring in Eqs. \eqref{eq:T_together} and \eqref{he.1}, by the corresponding expressions of a passive suspensions having the same density and temperature.
 Thus, in the limit of small spatial variations, i.e. in weakly inhomogeneous active liquids, they are represented in terms of gradients of the fields $n$, $T$ and ${\uu}$. 
The contributions of pressure tensor and heat flux vector stemming from the direct interactions among the particles, i.e. the ones not explicitly involving the self-propulsion, can be expressed using the standard macroscopic Navier expressions for the heat flux and momentum flux in terms of the hydrostatic pressure, the velocity gradients and the temperature gradient. 
This is because, at variance with previous approaches, the effect of the active force is not simply recast onto an additional stress contribution (leading to the well-known swim/active pressure \cite{takatori2014swim, winkler2015virial, levis2017active}), but its dynamics are still explicitly considered on the same footing as those of velocity and kinetic temperature fields \cite{epstein2019statistical}.
Hence, in the case of small velocity and temperature gradients, we write:
\begin{flalign}
\label{eq:q_gradT}
&{\bf q}=-\kappa \nabla T\\
&{\bf P}=\left( P_{h}-\left(\zeta-\frac{2}{3}\eta\right)\nabla\cdot \uu \right) {\bf I} - \eta 
\left( \nabla {\bf u}+\nabla {\bf u}^T \right) \,,
\label{eq:Pij}
\end{flalign}
where $\kappa$ is the thermal conductivity, $\eta$ the dynamic viscosity, $\zeta$ the so-called bulk viscosity and the superscript $T$ denotes the transpose of vectors or matrices.
The symbol $P_h$ represents the hydrostatic contribution to the pressure and, in equilibrium systems, it can be clearly identified in terms
of an ensemble average. In the present case, it can be obtained in principle by evaluating the trace of ${\bf P}^{(K)}+{\bf P}^{(C)}$
when the temperature and the velocity fields are uniform. For our scopes, we approximate such a quantity by using 
the hard-disks equation of state of Henderson~\cite{henderson1975simple}:
\begin{equation}
P_h(\mathbf{r})=n T \Bigl( \frac{1+y^2/8}{(1-y)^2}\Bigr)\,,
\end{equation}
where the effective temperature $T^*$ includes the effect of the active force and is defined later in Eq.~\eqref{eq:effective_temperatureb} (see the uniform solution reported in Sec.~\ref{eq:app_uniform}).
The remaining contributions to the pressure and heat flux involve
 the transport coefficients, $\kappa,\zeta,\eta$.
 which were fixed according to the Enskog theory of hard-disks. In the following, it was practical to replace them with
 the shear kinematic viscosity, the  longitudinal kinematic viscosity and the thermal diffusion coefficients defined as:
\begin{subequations}
\begin{align}
&
\nu_{\parallel}=\frac{1}{mn_0 }\left( \frac{4}{3}\eta+\zeta\right)\\
&
\nu_{\perp}=\frac{1}{mn_0 } \eta\\
&
D_T=\frac{2}{n_0 }  \kappa \,.
\end{align}
\end{subequations}
After using the expressions~\eqref{eq:q_gradT} and~\eqref{eq:Pij}),
the hydrodynamic equations turn out to be:
\begin{widetext}
\begin{subequations}
\begin{align}
&
\partial_{t}n = - \nabla\cdot( {\bf u}\,n) \label{v61_second}\\
&
\partial_t T + {\bf u}\cdot\nabla T+\frac{2 m}{n d}\gamma v_0  \uu \cdot{\bf p}
= \frac{2 }{nd}(\kappa\nabla^2 {\bf T} -  {\bf P}:\nabla{\bf u}) + 2\gamma(T_0 - T)+ \frac{2\gamma}{d}
m v_0   \langle \hn \cdot\vv\rangle_r \label{eq:T_together_second}\\
&
\partial_{t}{\bf u} + {\bf u}\cdot\nabla{\bf u} = - \frac{1}{m n}\left[\nabla P_h - \left(\nu_\parallel-\nu_{\perp}\right) \nabla (\nabla\cdot {\bf u}) -\nu_{\perp} \nabla^2 {\bf u} \right] - \gamma{\bf u} + \frac{{\bf f}^{ex}}{m}+ \gamma v_0\frac{{\bf p}}{n} \label{he.1_second}\\ 
&
\partial_{t} {\bf p} + \nabla\cdot  (n \langle \vv \otimes \hn \rangle_r)=-\frac{{\bf p}}{\tau} \label{he.1_polarization_f} 
\end{align}
\end{subequations}
\end{widetext}
where for compactness in Eq.~\eqref{eq:T_together_second} we employed ${\bf P}$ given by~\eqref{eq:Pij} .  
In this way, after eliminating $\langle \hn \cdot \vv\rangle_r$ and the elements of the tensor $\langle \hn \otimes \vv\rangle_r$ with the help of Eq.~\eqref{rlm2_2bis}, the hydrodynamic equations~\eqref{v61_second}-\eqref{he.1_polarization_f} are self-consistent and depend solely on the number, density, temperature, velocity and polarization densities fields. 
As  in the case of the Hydrodynamics of passive liquids, Eqs.~\eqref{v61_second}-\eqref{he.1_polarization_f} are non-linear and apart from a few exceptions, their solutions require numerical tools. For this reason, to obtain some insight, we shall look at their linearized version around a configuration describing a uniform active liquid.

\section{Uniform steady solutions of the transport equation}\label{eq:app_uniform}

The solution of the coupled system of hydrodynamic equations can only be achieved numerically, however, a useful starting  point, particularly relevant for our successive linear study, is to consider the stationary homogeneous state, where the system is subject only to a uniform driving force field
\begin{equation}\fvec^{ex}=m \gamma \uu \,,
\label{eq:fex_u_uniform}
\end{equation}
having introduced the uniform velocity $\uu$ to ease the notation.
Linearizing around this state consists of restricting the validity of our theory to homogeneous active liquids, excluding MIPS from the theoretical analysis.
The corresponding  FPE for the probability $\bar{f}=\bar{f}(\hn,\vv,t)$ formally coincides with Eq.~\eqref{many4}, upon replacing $\fvec^{ex}$ with Eq.~\eqref{eq:fex_u_uniform} and suppressing the spatial gradient terms in view of our hypothesis of spatial uniformity:
 \begin{eqnarray}\label{uniform}
 \frac{\partial}{\partial t} \bar{f} +
\gamma v_0 \hn \cdot   \frac{\partial }{\partial \vv}  \bar{f} && = \mathcal{L}_a
\bar{f} +
\\
&&
+\gamma  \frac{\partial }{\partial \vv} \Bigr[ \frac{T_0}{m}  \frac{\partial }{\partial \vv} +(\vv -\uu)\Bigr] \bar{f} +
\bar{\Omega} \nonumber\,,
\end{eqnarray}
where $\bar{\Omega}=\bar{\Omega}(\hn,\vv,t)$ and $\mathcal{L}_a$ is still given by Eq.~\eqref{eq:ActiveForce_evolutiongeneral}.
Notice that the coupling term in the l.h.s. between $\hn$ and $\vv$ is not symmetric.
This setup corresponds to a system of active particles (ABP or AOUP) subject to a constant force field which induces
a translation of the center of mass of the fluid at a constant velocity $\uu$. 

When $\Omega=0$,  i.e. when the interaction is suppressed,
the exact steady-state solution of Eq.~\eqref{uniform} can be obtained in the case of AOUP and reads:
\begin{equation}f_u(\hn,\vv)={\cal N} e^{-\frac{\beta_{vv}}{2} (\vv-\uu)^2-\frac{\beta_{ee}}{2}  \hn^2-\beta_{ev} (\vv-\uu)\cdot \hn}
\label{uniformsolution}
\end{equation}where ${\cal N} $ is a normalization factor and
\begin{eqnarray}&&
\beta_{vv}= \frac{1}{|\Sigma|} \nonumber
\\&&
\beta_{ee}= \frac{1}{|\Sigma|} \frac{T_0}{m}\left(1+\frac{m v_0^2}{T_0}\frac{\tau \gamma}{1+ \tau \gamma}   \right) \nonumber
\\&&
\beta_{ev}=-\frac{1}{|\Sigma|}\left( v_0\frac{\tau \gamma}{1+ \tau \gamma}    \right) \nonumber
\end{eqnarray}
with
\begin{equation}
|\Sigma|\equiv  \frac{T_0}{m}   \left( 1+\frac{m v_0^2}{T_0}\frac{\tau \gamma}{(1+ \tau \gamma)^2}   \right)  \,.   
\nonumber 
\end{equation}
This distribution formally corresponds to that of a free active particle (See, here for a discussion~\cite{caprini2021inertial}),
and is characterized by the following expectation values:
\begin{subequations}
\begin{align}
&
\label{eq:corr_matrx_e}
\langle \hn\rangle=0 \\
&
\label{eq:corr_matrx_v}
\uu= \frac{\fvec^{ex}}{ m \gamma}  \\
&
\label{eq:corr_matrx_ee}
\langle \hn \otimes \hn\rangle={\bf I} \\
&
\label{eq:corr_matrx_ev}
\langle  \hn \otimes (\vv-\uu)\rangle
= v_0\frac{\tau \gamma}{1+ \tau \gamma} {\bf I}  \\
&
\label{eq:corr_matrx_vv}
\frac{m}{d}\langle  (\vv-\uu)^2\rangle = T_0+ \frac{m v_0}{d} \langle \hn \cdot (\mathbf{v}-\mathbf{u}) \rangle \,.
\end{align}
\end{subequations}
Although it is not possible to write the exact uniform solution under the same form as Eq.~\eqref{uniformsolution}
 in the case of interacting particles, 
one may insist on looking for it as a multivariate Gaussian distribution. 
This idea is supported by particle-based numerical studies of both ABP and AOUP at a high density which reveal the almost Gaussianity of the single-particle velocity distribution also in regimes of large persistence~\cite{caprini2020active}. 
In practice, in the homogeneous high-density configurations, the influence of the active force could be recast onto a renormalization of the swim velocity, $v_0\to v[n]$, affecting the work performed by the active force.
In this way, we can include the interaction just by modifying the coefficients of the correlation matrix, i.e. Eq.~\eqref{eq:corr_matrx_ev} and Eq.~\eqref{eq:corr_matrx_vv} but not Eq. \eqref{eq:corr_matrx_ee} (let us observe that the latter is consistent with the closure employed for $\langle \hat{\mathbf{e}} \otimes \hat{\mathbf{e}} \rangle_r$).
This agrees with the assumption of Cates and Tailleur employed so far and means that Eqs.~\eqref{eq:corr_matrx_ev} and~\eqref{eq:corr_matrx_vv} should be replaced by:
\begin{subequations}
\begin{align}
&
\label{eq:corr_matrx_ev_interact}
\langle  \hn\otimes (\vv-\uu)\rangle= v[n] \frac{\tau \gamma}{1+ \tau \gamma} {\bf I}\\
&
\label{eq:corr_matrx_vv_interact}
\langle  (\vv-\uu)^2\rangle =d \frac{T_0}{m}\left( 1+\frac{m }{T_0}\frac{\tau \gamma}{1+ \tau \gamma}  v_0  v[n] \right)
\end{align}
\end{subequations}
The expectation values derived in this section 
(Eqs.~\eqref{eq:corr_matrx_e},~\eqref{eq:corr_matrx_v},~\eqref{eq:corr_matrx_ee} and Eqs.~\eqref{eq:corr_matrx_vv_interact} and~\eqref{eq:corr_matrx_vv_interact},  taking $\mathbf{f}^{ex}=0$) 
characterize the homogenous steady-state around which we linearize the hydrodynamic equations.

\section{Linearized Hydrodynamics}\label{Sec:linearizedhydro}

The theory  formulated by Landau and Lifshitz in 1957 allows studying the fluctuations of the hydrodynamic fields
around their averages~\cite{landau1959fluid}. It consists of linearizing the equations for the fields about their bulk values and study their small deviations when stochastic fluxes to the stress tensor and heat flux are added. The amplitudes of the noise terms are determined by the temperature and the transport coefficients of the fluid. In the following, for simplicity, we do not consider the contribution from the stochastic heat flux, but instead, include a stochastic polarization flux to capture the fluctuations of the active force.

First, we resort to a linearization scheme to simplify the structure of Eq.~Eq.\eqref{rlm2_2bis} and Eqs.~\eqref{v61_second}-\eqref{he.1_polarization_f}.
The linearization is performed around the stationary \emph{homogeneous} state, where the hydrodynamic fields (according to the results of Sec.~\ref{eq:app_uniform})  take the values $n=n_0$, $T=T^*$ , ${\bf u}=0$ and ${\bf p}=0$, 
where we have defined the effective temperature, $T^*$, for notational convenience:
 \begin{equation}
 T^*=T_0+m  \frac{\tau\gamma}{1+\tau\gamma} v_0  v[n]\,.
\label{eq:effective_temperatureb}
\end{equation}We stress that $T^*$ takes contributions from both the thermal agitation induced by the surrounding fluid bath and both the active force~\cite{petrelli2020effective}. As discussed before, the latter includes a phenomenological density-dependence~\cite{cates2015motility}.
Since in active colloidal and bacterial suspensions, $T_0\ll m v_0^2$~\cite{bechinger2016active} the effective temperature is usually dominated by the active contribution. 
 However, it is useful to keep the thermal contribution, since it yields the correct passive limit when $v_0 \to 0$.
We remark that our choice of the steady-state and $T^*$ is not arbitrary but, in the case of isotropic homogeneous active fluids,
follows from the assumptions made.

 To proceed further, we linearize the term $\langle \hn \otimes \vv\rangle_r$ in Eq.~\eqref{rlm2_2bis} around the uniform steady-state (Eqs.~\eqref{eq:corr_matrx_e},~\eqref{eq:corr_matrx_v},~\eqref{eq:corr_matrx_ee} and~\eqref{eq:corr_matrx_vv_interact}.
Keeping only terms linear in the fluctuations, Eq.~\eqref{rlm2_2bis} is approximated as:
\begin{eqnarray}\langle  \hn\otimes\vv\rangle_r \approx  && \frac{\tau \gamma  }{1+ \tau \gamma}\, v[n]  \Bigl[ {\bf I}  - \frac{\tau} {(1+ \tau\gamma)}\left(   {\bf I}\,\nabla\cdot\uu   + \nabla \otimes\uu  \right)\Bigr] \nonumber\\
&& -\frac{\tau}{1+\tau\gamma} \frac{ T^*}{ n_0 m }  \nabla \otimes {\bf p} \,.
\label{eq:nv_finalapprox2}
\end{eqnarray}
The structure of the hydrodynamic equations suggests separating the contribution of transverse and longitudinal components both for velocity and polarization fields, namely $\mathbf{u}_\perp$, $\mathbf{u}_\parallel$ and $\mathbf{p}_\perp$, $\mathbf{p}_\parallel$, to take advantage of the decomposition into curl-free and  divergence-free components.
In this way, the linearization about the steady-state (using also Eq.~\eqref{eq:nv_finalapprox2}) leads to a couple of equations for $\mathbf{u}_\perp$ and $\mathbf{p}_\perp$, that are not affected by the other fields:
\begin{subequations}
\begin{align}
&
\label{eq:p_longitudinal}
\partial_{t}{\bf u_\perp}  =  
 \nu_{\perp}\nabla^2{\bf u_\perp} - \gamma{\bf
  u_\perp} +\frac{1}{n_0}\gamma v_0 {\bf p_\perp}\\ 
&
\frac{\partial}{\partial t}  {\bf p_\perp}=
\frac{\tau\gamma}{1+\tau\gamma}
\Bigl[\frac{ T^*}{m \gamma} \nabla^2 {\bf p_\perp} \Bigr]
 -\frac{1}{\tau}   {\bf p_\perp} \,. 
\label{eq:p_trasversal} 
\end{align}
\end{subequations}
The form of the equations for the longitudinal fields suggests defining the following quantities to simplify the calculations:
\begin{subequations}
\label{campipsiphi}
\begin{align}
&\psi\equiv \nabla
\cdot{\bf u_\parallel} \,,
\\
&\phi\equiv \nabla
\cdot{\bf p_{\parallel}} \,.
\end{align}
\end{subequations}
%
\begin{widetext}
Taking the divergence  of Eqs.~\eqref{he.1_second} and~\eqref{he.1_polarization_f} and linearizing about the homogeneous state (using also Eq. \eqref{eq:nv_finalapprox2}), in the case of a two-dimensional system ($d=2$), 
we obtain the following set of four coupled equations for $n$, $T$, $\psi$, $\phi$:
\begin{subequations}
\begin{align}
&
\label{eq:lin_density}
\partial_{t}n= - n_0\psi
\\
&
\label{eq:lin_temperature}
\partial_t T = \frac{1 }{n_0 }\left(\kappa\nabla^2T- P_h\psi \right)+2\gamma \left( (T^*-T) -\frac{m v_0}{2} \frac{\tau\gamma}{1+\tau\gamma}\left(2 v_0  \frac{\delta n}{n_c} +\frac{ T^*}{m n_0\gamma} \phi \right) \right) -   \frac{ 3 m v_0 v[n] \tau^2 \gamma^2}{(1+ \tau \gamma)}\psi
 \\
&
\label{eq:lin_velocitylongitudinal}
 \partial_{t}\psi  = -   \frac{v_s^2}{n_0}\nabla^2 n - \alpha v^2_s  \nabla^2 T+ \nu_{\parallel}\nabla^2\psi - \gamma \psi +\frac{1}{n_0}\gamma v_0 \phi\\ 
&
\label{eq:lin_polarizationlongitudinal}
\frac{\partial}{\partial t}  \phi=- \frac{\tau\gamma}{1+\tau\gamma}\left( \left(v[n] - v_0\frac{n_0}{n_c}  \right) \nabla^2 n- \frac{ T^*}{m \gamma} \nabla^2 \phi -2\frac{\tau }{1+\tau\gamma}  n_0  v[n]  \nabla^2 \psi   \right)-\frac{1}{\tau}\phi  \,,
\end{align}
\end{subequations}
\end{widetext}
where we have introduced the sound speed, $v_s$, 
\begin{equation}
v_s^2=\frac{1}{m} \left[ \frac{\partial P_h}{\partial n} \right]_T \,,
\end{equation}
 the thermal expansion coefficient $\alpha$,
\begin{equation}
\alpha=-\frac{1}{n_0} \left[ \frac{\partial n}{\partial T} \right]_P \,,
\end{equation}
and used the following thermodynamic identity:
\begin{equation}\frac{1}{mn_0}\left[ \frac{\partial P_h}{\partial T} \right]_V =\alpha\,  v_s^2 \,.
\end{equation}
It is convenient to evaluate the two sets of Eqs.~\eqref{eq:p_longitudinal}-\eqref{eq:p_trasversal} and Eqs.~\eqref{eq:lin_density}-\eqref{eq:lin_polarizationlongitudinal} in the Fourier reciprocal space. 
The solutions depend on the phenomenological parameter $n_c$, and the transport coefficients, such as the transverse and longitudinal viscosities $\nu_{\perp}$ and $\nu_\parallel$, and the sound speed $v_s^2$. Additional parameters, such as the thermal expansion coefficient $\alpha$ and the thermal conductivity $\kappa$, appear because of the coupling between velocity and temperature fields.
Since, to the best of our knowledge, there are no analytical expressions for the transport coefficients of active liquids,
we resorted to a drastic but clear choice for their estimate.
To fix all these quantities, we employed one of the theories of major success dealing with the hard-disk systems assimilating the steric effects of the interacting active particles to those of two-dimensional disks and treated the contribution of the self-propulsion as an additive effect. 
Although this choice is somehow arbitrary, it should give reasonable information about the relative importance of the transport coefficients in the expressions for the correlations.
As shown in Appendix~\ref{app:Parameter}, these parameters contain the activity only via $T^*$ (Eq.~\eqref{eq:effective_temperatureb}), so that they show a strong dependence on $v_0$, while $\tau$ and $\gamma$  play a marginal role. 
For high density, we expect that the thermal expansion coefficient, $\alpha$, remains small in such a way that, in Eq.~\eqref{eq:lin_velocitylongitudinal}, the coupling term between velocity and temperature fields, that is $\sim \alpha v_s^2$ is weak. 
The argument sketched here is described in detail in Appendix~\ref{app:demonstration} where the explicit expression of $\alpha$ is reported.
With this simplification which leads to more clear and transparent results, we assume that the temperature field effectively decouples from the remaining longitudinal modes so that the $4\times4$ linear system given by Eqs.~\eqref{eq:lin_density}-\eqref{eq:lin_polarizationlongitudinal}, is reduced to a $3\times3$ problem, namely Eqs.~\eqref{eq:lin_density},~\eqref{eq:lin_velocitylongitudinal} and~\eqref{eq:lin_polarizationlongitudinal} with $\alpha=0$.
\section{Fluctuating Hydrodynamics}\label{Sec:fluctuatinghydro}
\label{fluctuatingHydrodynamics}

To solve the linear system of partial differential equations for the deviations $\delta a=a-\overline{a}$ of the hydrodynamic fields
from their homogeneous value, $\overline{a}$,
we consider their  Fourier transforms (denoted by hat-symbols):
\begin{equation}
\delta\hat{ {\bf a}}({\bf k},t)=\int d{\bf r}~\delta {\bf a}({\bf r},t)e^{- i{\bf k}\cdot{\bf r}} \,.
\label{ft.1}
\end{equation} 
where we used
the following compact notation:
\begin{equation}
\delta \hat {{\bf a}}({\bf  k},t)=\{\delta \hat{n}({\bf k},t), \hat{\psi}({\bf  k},t),  \hat{\phi}({\bf k},t),  \hat{u}_\perp({\bf k},t), \hat{p}_\perp({\bf k},t)\} \nonumber\, .
\end{equation}
The transverse modes of velocity and polarization fields read: 
\begin{subequations}
\begin{align}
\hat {u}_\perp({\bf k}) &= \hat{k}_\perp \cdot \hat{{\bf u}}({\bf k})\,, \\
\hat {p}_\perp({\bf k}) &= \hat{k}_\perp \cdot \hat{ {\bf p}}({\bf k}) \,, 
\end{align}
\end{subequations}
where $\hat{k}_\perp$ is a unit vector such that $\hat{k}_\perp \cdot\hat{k}=0$.  
Following the methods of the fluctuating Hydrodynamics, we study the fluctuations of the hydrodynamic fields
by considering the stochastic equations obtained by adding suitable noise sources to
 Eqs.~\eqref{eq:p_longitudinal}-\eqref{eq:p_trasversal} and Eqs.~\eqref{eq:lin_density}-\eqref{eq:lin_polarizationlongitudinal}.
 These equations in Fourier representation read:
\begin{equation}  
\frac{d}{dt }\delta\hat{\bf a}({\bf k},t)=-{\bf M}(k)\cdot\delta \hat{{\bf a}}({\bf k},t)+ \hat{\bfxi}({\bf k},t) \,,
\label{akequation}
\end{equation}
where the matrix ${\bf M}(k)$ is made up of two blocks and has the following representation:
\begin{widetext}
\begin{equation}
\hspace{-2.5cm}\mathbf{M} = 
\left( \begin{array}{cccccc}
0 & n_0 & 0&0&0\\ 
 - \frac{v_s^2}{n_0} k^2  &\gamma+  \nu_{\parallel} k^2  &  -\frac{1}{n_0}\gamma v_0&0&0  \\ 
- \frac{\tau\gamma}{1+\tau\gamma}\Bigl(v[n] - v_0 \frac{n_0}{n_c} \Bigr) k^2  &  \frac{2}{\gamma}(\frac{\tau\gamma}{1+\tau\gamma} )^2  n_0  v[n] k^2 &\frac{1}{\tau}+ \frac{ T^*}{m \gamma} \frac{\tau\gamma}{1+\tau\gamma} k^2&0&0\\ 
0&0&0&\gamma+  \nu_{\perp} k^2 &-\frac{1}{n_0}\gamma v_0 \\ 
0&0&0&0&\frac{1}{\tau}+ \frac{ T^*}{m \gamma} \frac{\tau\gamma}{1+\tau\gamma} k^2 
\end{array} \right) \, .
\label{fh.1}
\end{equation}
\end{widetext}
The term $\hat{\bfxi}({\bf k},t)$ is a noise field that models the contribution of both the stochastic thermal forces due to the interaction with the solvent and the fast degrees of freedom (viz. the higher moments)
which are not accounted for in the hydrodynamic description. 
We fix the noise by requiring that, when the active force vanishes ($v_0=0$), the steady-state velocity fluctuations reduce to those corresponding to an equilibrium system. 
In other words, we assume that the fluctuation-dissipation theorem holds in the reference passive system.
The derivation of this principle is described in detail in Appendix~\ref{app:noise_amplitude} both for the longitudinal and the transverse fields. 
According to this principle the noise has the following properties:
\begin{eqnarray}&&
\langle\hat{\boldsymbol{\xi}}(t, \mathbf{k}) \rangle=0 \nonumber\\&&
\langle \hat{\boldsymbol{\xi}}(t, \mathbf{k}) \otimes\hat{\boldsymbol{\xi}}(t', -\mathbf{k}) \rangle=2 {\bf D}(\mathbf{k}) \delta(t-t') \nonumber \,,
\end{eqnarray}
where $\hat{\boldsymbol{\xi}}$ has the same number of components as $\hat{\mathbf{a}}$ and the effective diffusion matrix
${\bf D}$ is diagonal with the following non-vanishing elements:
\begin{subequations}
\begin{align}
&
 D_{\psi\psi}=\frac{ T_0}{m} k^2( \gamma+  \nu_{\parallel} k^2 ) \\
&
 D_{\phi\phi}= n_0^2\Bigl(\frac{1}{\tau}+ \frac{ T^*}{m \gamma} \frac{\tau\gamma}{1+\tau\gamma} k^2\Bigr)  \frac{v[n]}{v_0}\,k^2\\
&
D^{\perp}_{uu}= \frac{ T_0}{m}  ( \gamma+  \nu_{\perp} k^2 )    \\
&
D^{\perp}_{pp}= n_0^2 \Bigl(\frac{1}{\tau}+ \frac{ T^*}{m \gamma} \frac{\tau\gamma}{1+\tau\gamma} k^2 \Bigr) \frac{v[n]}{v_0} \,.
\end{align}
\end{subequations}
The validity of the fluctuation-dissipation theorem only constrains the noise correlations in the equilibrium limit $v_0 \to 0$. 
 In principle, in the active case, the hydrodynamic noise could have different expressions for the diffusion matrix and even non-white time-correlations (for instance if a strong separation of scale is lacking, between fast and slow variables). 
A systematic coarse-graining procedure that derives rigorously the properties of the noise starting from the full Fokker-Planck equation of the system (Eq.~\eqref{many4}) is typically very difficult and certainly beyond our scope. 
Historical examples have been obtained in the case of molecular fluids~\cite{zwa,hinton,fox}, granular fluids~\cite{brey},
and lattice models~\cite{manac1,manac2}.

\section{Equal-time correlations}
\label{equaltimecorrelations}
After fixing the diffusion matrix with elements $D_{\alpha \beta}$, the equal-time (stationary) correlations in Fourier space, defined as 
\begin{equation} 
{\bf C}(k)=\langle \delta \hat {\bf a}(\mathbf{k}) \otimes\delta\hat {\bf a}(-\mathbf{k})\rangle \,,
 \end{equation}
may be determined by solving the Lyapunov equations associated with the dynamics~\eqref{akequation}: 
\begin{equation}
{\bf M}(k) \cdot {\bf C}(k)+{\bf C}(k) \cdot {\bf M}^{T}(-k)=2 {\bf D} \,.
\label{structurefactors}
\end{equation}
Hereafter, we restrict to the velocity-velocity spatial correlation evaluating separately transverse and longitudinal components as suggested by the block-structure of the matrix $\mathbf{M}$.

\subsection{Transverse equal-time  correlations}

Considering the lower-right block of the matrix~\eqref{fh.1}, ${\bf M}^\perp$, 
we determine the Fourier transform of the equal-time velocity-velocity transverse correlations, $C^\perp({\bf k})$,  defined as:
\begin{equation}
C^\perp({\bf k})=\langle |\hat{\mathbf{u}}_\perp(\mathbf{k})|^2\rangle \, .
\end{equation}
For the sake of conciseness, in the following we shall denote by $\perp$ and $\parallel$ the transverse and longitudinal elements of the tensors.
The details of the calculations are reported in Appendix~\ref{app:noise_amplitude} and lead to the formula:  
\begin{equation}
C^\perp({\bf k}) \approx \frac{T_0}{m} + \frac{  v_0 v[n]  \tau\gamma}{1+{\tau\gamma }}\frac{1}{1+\xi^2_\perp k^2} \,,
\label{transversecorrelation}
\end{equation}%
where $\xi_\perp$ is the correlation length associated with the transverse mode of the spatial velocity correlation, which reads: 
\begin{equation}\xi^2_\perp=\frac{\tau \gamma}{1+\tau \gamma}
\left[ \frac{\tau}{\gamma}\frac{1}{1+\tau\gamma} \frac{ T^*}{m }+    \frac{\nu_\perp}{\gamma}\left(2+\frac{1}{\tau\gamma}\right)\right] \,.
\label{eq:xi_perp}
\end{equation}$C^\perp(\mathbf{k})$ displays an Ornstein-Zernike form, viz. decays exponentially in real space, a property not having an equilibrium counterpart.
Indeed, in the thermal limit, $v_0\to 0$ or $\tau \to 0$, the $k$-dependence in Eq.~\eqref{transversecorrelation}
 disappears and the amplitude of $C^\perp(\mathbf{k})$ is simply $T_0/m$, as expected at equilibrium.
When $\gamma\tau\ll1$, inertial effects induce a similar suppression of the spatial ordering.
The spatial velocity correlations become irrelevant when the so-called active temperature is smaller than the solvent temperature, i.e. $v_0^2 \tau\gamma \ll T_0/m$  (see Eq.~\eqref{transversecorrelation}).
The expression of $\xi_\perp$ contains two distinct contributions: i) the ``thermal'' one, which depends on $T^*$ and, thus, both on swim velocity and solvent temperature (usually $v_0^2 \gg T_0/m$ in active colloids);
ii) the second one proportional to the transverse viscosity $\nu_\perp$. 
According to the estimates of Appendix~\ref{app:Parameter}, the dependence on $\tau$ appears only explicitly both in Eq.~\eqref{transversecorrelation} and Eq.~\eqref{eq:xi_perp} because $\nu_\perp$ and $T^*$ are $\tau$-independent:

In the overdamped limit, $\tau\gamma\gg1$, $\xi_\perp$ does not vanish showing that the
presence of a finite correlation length is not a consequence of the inertial dynamics. Specifically:
\begin{equation}\lim_{\gamma\tau\gg1}\xi^2_\perp \approx \frac{v_0 v[n] }{\gamma^2}  + 2\frac{\nu_\perp}{\gamma}   \,,
\label{eq:xi_perp_over}
\end{equation}where we used Eq.~\eqref{eq:effective_temperatureb} to eliminate the temperature $T^*$ and neglected the small contribution $T_0$.
In the dense regime, we expect that the term proportional to $ \nu_{\perp}$ dominates because the viscosity grows with the packing faster than the first term.
Equation~\eqref{eq:xi_perp_over} has an explicit decreasing dependence on $\gamma$: the larger the friction, the smaller $\xi_\perp$.
Interestingly, $ \xi_{\perp}$ does not depend on $\tau$ ($\nu_\perp$ is $\tau$-independent), a prediction almost in agreement with recent observations made by Szamel et al. who numerically studied the transverse correlation length of active liquids in particle-based simulations (in the overdamped regime)~\cite{szamel2021long}.

It is instructive to evaluate Eq.~\eqref{eq:xi_perp} in the opposite inertial limit, $\tau \gamma \ll 1$, assuming that the active temperature remains larger than the thermal temperature.
In this case, the expression for $\xi_\perp$ reads:
\begin{equation}
\lim_{\gamma\tau\ll1} \xi^2_\perp \approx   \tau^2 v_0  v[n] + \frac{\nu_\perp}{\gamma}   \,.
\label{eq:xi_perp_under}
\end{equation}
In this limit,  the term $\propto \nu_\perp$ in the expression of $\xi_\perp$ is still $\tau$-independent while the first term scales as $\sim\tau$. 
 Again, one could expect that the $\nu_\perp$ term is dominant because of its dependence on the packing fraction although particle-based simulations have yet to be performed, in this case.

\subsection{Longitudinal equal-time correlations}

The longitudinal spatial velocity correlation, defined by
\begin{equation}
C^\parallel(\mathbf{k}) = \langle |\hat{\mathbf{u}}_\parallel(\mathbf{k})|^2\rangle\,,
\end{equation} 
is obtained by solving Eq.~\eqref{structurefactors} considering the reduced $3\times3$ problem, employing the matrix ${\bf M}^\parallel(\mathbf{k})$ given by the upper left block of the expression~\eqref{fh.1}.
Going back from $\psi$ to $\uu_\perp$,
we obtain (see Appendix~\ref{app:noise_amplitude}):
\begin{equation}
C^\parallel(\mathbf{k}) =\frac{T_0}{m}   +   \frac{  v_0 v[n]}{1+\frac{1}{\tau \gamma}}\frac{1}{1+\xi^2_\parallel k^2}
 \label{velocityfluctuation}
  \end{equation}
where $\xi_\parallel$ is the correlation length associated with the longitudinal modes of the velocity field, that reads:
 \begin{equation}
\begin{aligned}
\xi_\parallel^2 =\frac{\tau\gamma}{1+\tau\gamma} &\Biggl[ \frac{ T^*}{m\gamma^2 }\frac{\tau \gamma}{\tau\gamma+1} +\frac{\nu_{\parallel}}{\gamma^2}  \left(2\gamma+\frac{1}{\tau}\right)\\
& +2  \frac{\tau}{\gamma} v_0 v[n] \left(\frac{\tau \gamma}{1+\tau\gamma} \right)^2     + \frac{\tau}{\gamma} v_s^2 \Biggr] \,.
\label{eq:xi_parallel}
\end{aligned}
\end{equation}
Likewise $C^\perp({\bf k})$, also $C^\parallel({\bf k})$ has an Ornstein-Zernike form corresponding to 
an exponential-like decay in real space and 
 becomes $k$-independent both in the thermal equilibrium limit, $v_0 \to 0$, and in the inertial limit, $\tau\gamma\ll1$.
The major difference between Eq.~\eqref{transversecorrelation} and Eq.~\eqref{velocityfluctuation} relies in the 
dependence of the two correlation lengths $\xi_\perp$ and $\xi_\parallel$ on the control parameters. 
The expression for $\xi_\parallel$ contains four terms:
the first three are similar to those of the expression for $\xi_\perp$ and depend on $T^*$, $v_0^2$ and the longitudinal kinematic viscosity, $\nu_\parallel$;
the last term is determined by the sound speed, $v_s$ which has been estimated in Appendix~\ref{app:Parameter} and that does not depend on $\tau$. 

In the overdamped limit, $\tau\gamma\gg1$, the longitudinal correlation length assumes a simpler form:
\begin{flalign}
\hspace{-0.15cm}\lim_{\gamma\tau\gg1}\xi_\parallel^2 &= \frac{\tau}{\gamma} \left(    v_s^2 +2   v_0 v[n] \right)  +2\frac{\nu_{\parallel}}{\gamma} +  \frac{ T^*}{m\gamma^2 }            
\approx \frac{\tau}{\gamma} v_s^2\,.
\label{eq:xi_parallel_over}
\end{flalign}
While all terms ( provided $v[n]>0$) represent positive contributions to $\xi_\parallel$, the $v_s^2$ term becomes dominant because it has the fastest increase when the packing fraction becomes large.
The coherence length, $\xi_\parallel$, displays a strong dependence on the persistence time, scaling as $\sim\sqrt{\tau}$,
a result in agreement with the numerical results by Szamel et al.~\cite{szamel2021long} based on particle simulations.
Interestingly, this prediction is also in accord with the microscopic theory developed for the case of active solids where $v_s^2$ is proportional to the second derivative of the interaction potential calculated at the lattice constant of the solid.
 
Finally, we discuss the expression for $\xi_\parallel$ in the underdamped regime, $\tau\gamma \ll 1$, always assuming that the active temperature is larger than the solvent temperature:
\begin{equation}
\lim_{\gamma\tau\ll1}\xi_\parallel^2 =  \frac{\nu_\parallel}{\gamma} + \tau^2 v_s^2     \approx \tau^2 v_s^2\,.
\label{eq:xi_parallel_under}
\end{equation}
 Equation~\eqref{eq:xi_parallel_under} shows that, in this case, $\xi_\parallel$ has a faster growth with $\tau$  
 than in the overdamped regime, where $\xi_\parallel\sim \sqrt{\tau}$.
The linear scaling obtained in the inertial regime agrees with the numerical and theoretical results obtained by particle-simulations in inertial active solids~\cite{caprini2021spatial} while further numerical studies are needed to check this result in inertial active liquids.

The above discussion shows that the main difference between liquid and solid (both in overdamped and underdamped regimes) is the presence of a much shorter velocity correlation length of the transverse modes with respect to the longitudinal length, $\xi_\parallel \ll \xi_\perp$. 
Indeed, the solid supports traveling waves of both transverse and longitudinal type, while through the bulk of a fluid (liquid or gas) only longitudinal waves can propagate. 
If the medium is not rigid (fluids), the particles will slide past each other and will not generate a transverse wave but only a diffusive momentum propagation (the shear diffusion mode).

\section{Dynamical correlations}
\label{dynamical}

The study of the time-dependent correlations reveals some new interesting aspects 
of the dynamics of the active system at hand.
The dynamical structure factors, defined as
\begin{equation}
{\bf S}(k,\omega)=\langle \delta \tilde {\bf a}(\mathbf{k}, \omega) \otimes\delta\tilde {\bf a}(-\mathbf{k}, \omega)\rangle
\end{equation}
 are obtained by solving Eq.~\eqref{akequation}, in the frequency domain:
\begin{equation}
\widetilde{{\bf M}}({\bf k},\omega)\cdot\delta\tilde{{\bf a}}({\bf k},\omega)=\tilde{\mbox{\boldmath $\xi$}}({\bf k},\omega) \,,
\label{dsf.2}
\end{equation}
with $\widetilde{{\bf M}}({\bf k},\omega)=i\omega {\bf I}+{\bf M}({\bf k})$. The vectors $\tilde{\mbox{\boldmath $\xi$}}({\bf k},\omega)$ and $\delta\tilde{{\bf a}}({\bf k},\omega)$ are the time-Fourier transform of $\hat{\bfxi}({\bf k},t)$ and $\delta\hat{\bf a}({\bf k},t)$, respectively. The latter is defined as:
\begin{equation}
\delta \tilde{{\bf a}}({\bf k},\omega)=\int_{-\infty}^{\infty} dt~\delta {\bf a}({\bf k},t)e^{-i\omega t} \,,
\end{equation}
where $\omega$ is the frequency and a similar definition holds for the vector of noise, $\tilde{\mbox{\boldmath $\xi$}}({\bf k},\omega)$,
characterized by zero average and the following correlations:
\begin{equation}
\langle\tilde{\mbox{\boldmath $\xi$}}({\bf k},\omega)\otimes\tilde{\mbox{\boldmath $\xi$}}(-{\bf k},\omega')\rangle
=2{\bf D}({\bf k})\delta(\omega+\omega') \,,
\label{dsf.4}
\end{equation}
with the matrix $\mathbf{D}$ introduced in Sec.~\ref{equaltimecorrelations}. 
Multiplying Eq.~\eqref{dsf.2} on the left by $\widetilde{{\bf M}}^{-1}(k,\omega)$ and on the right by $\delta\tilde{{\bf a}}^{T}(-k,-\omega)$  
and averaging over the noise, we obtain the matrix of dynamical structure factors:
\begin{equation}
\begin{aligned}
{\bf S}(k,\omega)&=
\langle\widetilde{{\bf M}}^{-1}(k,\omega)\cdot\left[\tilde{\mbox{\boldmath $\xi$}}(k)\otimes\delta\tilde{{\bf a}}^{T}(-k,-\omega)\right]\rangle \\
&=2\,\widetilde{{\bf M}}^{-1}(k,\omega) \cdot {\bf D}(k)\cdot
[\widetilde{{\bf M}}^{T}(-k,-\omega)]^{-1} \,,
\label{dsf.5}
\end{aligned}
\end{equation}
where in the last equality we have used the Hermitian conjugate of
Eq.~\eqref{dsf.2} and the relation~\eqref{dsf.4}.
Back-transforming from $\omega$ to $t$, we obtain the two-time correlation structure factors in the 
$(k,t)$ representation, namely the matrix of intermediate scattering functions ${\bf F}(k,t)$.

\subsection{Hydrodynamic spectrum}

\begin{figure}[t]
\begin{center}
\includegraphics[width=0.95\linewidth,keepaspectratio,angle=0,clip=true]{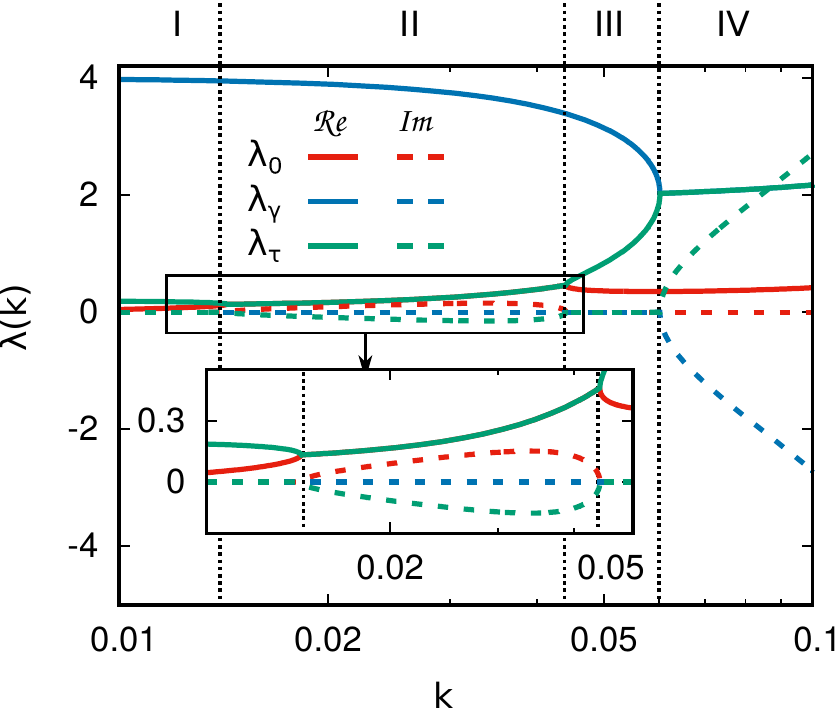}
\caption{Eigenvalues of the dynamical matrix $M$ versus wavenumber $k$ (to get an idea of lengthscales, we recall that the effective diameter of particles is set to $\sigma=1$). We also recall that positive real parts indicate fluctuations that decay in time, i.e. stable modes while imaginary parts denotes the presence of propagating waves.
As discussed in the text, four different regions, marked by vertical black dotted lines, are identified. 
At very small $k$ (region I) all eigenvalues are real and they tend, for $k \to 0$, to $0,\gamma,1/\tau$ (for $\lambda_0$,$\lambda_\gamma$ and $\lambda_\tau$ respectively). 
The values of the parameters are: $n_0=0.7$, $n_c=2$, $\gamma=4$, $\tau=5$, $m=1$, $d=2$, $v_0=10$, $T_0=0.1$.
The analytical prediction Eqs.~\eqref{lambda_n}-\eqref{lambda_phi} holds only in region I.
}
\label{FitDinamici}
\end{center}
\end{figure}


 Before delving into the discussion of the dynamical structure factor, we consider the dependence of the eigenvalues of $\mathbf{M}$ as a function of the wavevector $k$ and the control parameters.
In simple liquids, each eigenvalue of the dynamical matrix is associated with a particular fluctuation such as a sound mode or a shear mode.
It might be tempting to extend such correspondence to the active case, including two additional eigenvalues associated with the fields ${\bf p}_\parallel$ and ${\bf p}_\perp$. 
 However, as the analysis (see Sec.\ref{struf}) of the intermediate scattering functions demonstrates, in active liquids these identifications become unclear, since the relaxation of each hydrodynamic fluctuation is determined by more than one eigenvalue.

We first obtain the eigenvalues associated with the transverse fluctuations from ${\bf M}^\perp$ (the lower right $2\times 2$ block of $\mathbf{M}$ in \eqref{fh.1}). They are real and positive for any choice of the parameters:
\begin{subequations}
\begin{align}
&
\lambda^{\perp}_\gamma(k)=\gamma+\nu_\perp k^2 \, .
\label{eq:autotu}\\
&
\lambda^{\perp}_\tau(k)=\frac{1}{\tau}+\frac{\tau\gamma}{1+\tau\gamma} \frac{ T^*}{m \gamma} k^2 \,.
\label{eq:autotp}
\end{align}
\end{subequations}
In passive liquids, the eigenvalue $\lambda^{\perp}_\gamma(k)$ describes shear waves and is associated with the diffusion of the component of the momentum orthogonal to the direction of a propagating signal.
At variance with simple (inviscid) liquids~\cite{hansen1990theory}, this eigenvalue does not vanish as $k\to 0$ due to the presence of the solvent drag force proportional to $\gamma$ and grows as $k^2$ with a coefficient given by the shear kinematic viscosity, $\nu_\perp$, that in 
the present treatment depends on
the active force through $T^*$.
However, the second eigenvalue, $\lambda^{\perp}_\tau(k)$, describes
a transverse polarization fluctuation that has not a passive counterpart. As we shall see below,
this eigenvalue is important to understanding how the velocity field relaxes.
We remark that the $k^2$-terms both in Eqs.~\eqref{eq:autotu} and~\eqref{eq:autotp} do not increase with $\tau$ in the overdamped regime but their amplitudes are mostly determined by $v_0$.
In the overdamped regime, we have the following inequality: $\lambda^{\perp}_\gamma(k)\gg \lambda^{\perp}_\tau(k)$ (while in the inertial regime the opposite relation holds), a property that will influence the dynamic properties. 

 We turn, now, to the longitudinal modes by solving the eigenvalue problem for ${\bf M}^\parallel$ (the upper left $3\times 3$ block of $\mathbf{M}$ in \eqref{fh.1}).
When $k=0$ and $v_0=0$ (passive limit), one can unambiguously identify a $\lambda_0$-mode which describes a density fluctuation,
and a $\lambda_\gamma$-mode with a momentum fluctuation.  
In the active case, an additional polarization fluctuation will be described by a third eigenvalue $\lambda_\tau$ without a passive counterpart which will be crucial also for the decay of velocity fluctuations. 
At first, we obtain perturbatively the longitudinal eigenvalues  by expressing them as an expansion in powers of $k$ around the $k=0$-values, namely $0,\gamma,1/\tau$ (for $\lambda_0$, $\lambda_\gamma$ and $\lambda_\tau$ respectively). 
Up to quadratic order, we find the following real and positive expressions:
\begin{widetext}
\begin{subequations}
\begin{align}
&
\label{lambda_n} 
\lambda_0\approx \Biggl[\frac{v_s^2}{\gamma} +
\frac{v_0}{\gamma}  \frac{\tau^2\gamma^2}{1+\tau\gamma}
 \left(v[n]    -  v_0\frac{n_0}{n_c} \right)\Biggr] k^2 \\
& 
\label{lambda_psi} 
 \lambda_\gamma\approx\gamma-\Biggl[\frac{v_s^2}{\gamma}- \nu_\parallel+\frac{v_0 }{\gamma-\frac{1}{\tau}}\Biggl[2 \left(\frac{\tau \gamma}{1+\tau\gamma} \right)^2   v[n]   
 -  \frac{\tau\gamma}{1+\tau\gamma}\left(v[n]    -  v_0\frac{n_0}{n_c}\right) \Biggr] k^2 \to \gamma -  \frac{v_s^2}{\gamma} k^2\\
&
\label{lambda_phi}
 \lambda_\tau\approx\frac{1}{\tau}+\Biggl[\frac{ T^*}{m \gamma} \frac{\tau\gamma}{1+\tau\gamma}+\frac{v_0}{  \left(\gamma-\frac{1}{\tau}\right)}\left(2 \left(\frac{\tau \gamma}{1+\tau\gamma} \right)^2   v[n]   -    \frac{\tau^2\gamma^2}{1+\tau\gamma}\left(v[n]    - v_0 \frac{n_0}{n_c}\right) \right) \Biggr]k^2  \to \frac{1}{\tau} - v_0 v[n] \tau^2 \gamma k^2\,,
\end{align}
\end{subequations}
\end{widetext}
where after the symbol $\to$ we have written the dominant contributions in the overdamped regime, $\tau\gamma \gg 1$.
The approximate expressions~\eqref{lambda_n},~\eqref{lambda_psi} and~\eqref{lambda_phi} are not valid for values of $k$ above a certain threshold where the eigenvalues become complex. 
To explore these regimes, Fig.~\ref{FitDinamici} shows the three eigenvalues as a function of $k$ for a particular (but relevant) choice of the control parameters, corresponding to an overdamped system ($\tau \gamma \gg 1$) with a large active force such that, $m v_0^2 \gg T_0$.
Bearing in mind that positive real parts provide stable modes, i.e. fluctuations that decay in time, while imaginary parts denote the presence of propagating waves, we identify four different regimes labeled with Roman numerals (we do not explore larger values of $k$ because our theory only applies to small gradients).
We employ three different colors to identify each eigenvalue and distinguish real and imaginary parts using solid and dashed lines, respectively.

\begin{figure}[!t]
\centering
\includegraphics[width=0.95\linewidth,keepaspectratio,angle=0]{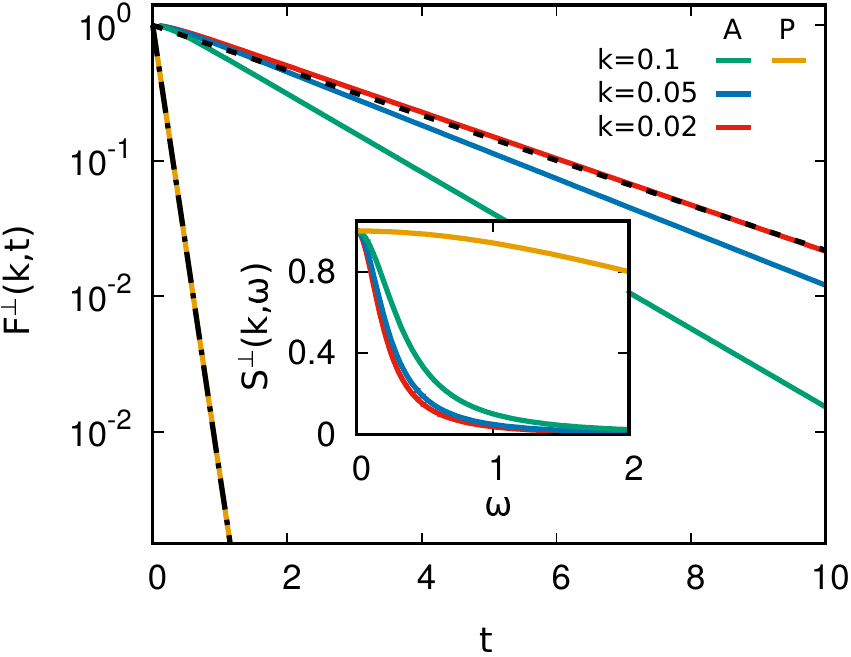}
\caption{\label{figintermediate} Dynamical properties of the transverse modes.
Main panel: transverse velocity intermediate scattering function $F^\perp(k,t)$ versus $t$ for three different values of $k$ for both for active (A) and passive (P) cases. 
The four solid curves are obtained by the exact formula~\eqref{e104b} and the two exponentials $e^{-\gamma t}$ (expected for passive liquids) and $e^{-t/\tau}$, are plotted as dashed and dotted-dashes lines. 
Inset: $S^\parallel(k,\omega)$ as a function of $\omega$ and fixed $k$ according to the legend of the main panel. 
Both $F^\perp(k,t)$ and $S^\parallel(k,\omega)$ has been normalized with their values at $t=0$.
The values of the parameters are the same as in Fig.~\ref{FitDinamici}.}
\end{figure}


For the smallest values of $k$ (region I), the longitudinal eigenvalues are real and positive and are roughly described by the predictions~\eqref{lambda_n},~\eqref{lambda_psi} and~\eqref{lambda_phi}.
Starting from their values at $k=0$ ($0,\gamma,1/\tau$), they remain distinct: while the $\lambda_0$-mode represents a diffusive mode (with diffusion constant $\approx v_s^2/\gamma $), both the fluctuations associated with $\lambda_\gamma$ and $\lambda_\tau$ decay in time
at a finite rate even when $k=0$. The eigenvalue
$\lambda_{\gamma}$ decreases with $k$  proportionally to $-v_s^2/\gamma$, while $\lambda_\tau$ decreases as $-v_0 v[n] \tau^2 \gamma k^2$
(see the inset of Fig.~\ref{FitDinamici}).
Remarkably, the active force affects the longitudinal eigenvalues through the $T^*$ dependence in the transport coefficients (as in the transverse case), but the expression for $\lambda_\tau$ decreases faster for increasing $\tau$ at variance with $\lambda^\perp_\tau$.
This difference reflects the one occurring between $\xi_\parallel$ and $\xi_\perp$ in the overdamped case (see Eq.~\eqref{eq:xi_parallel_over} and Eq.~\eqref{eq:xi_perp_over}).

Region II is different because the $\lambda_0$ and $\lambda_\tau$ modes (red and green curves) mix and give rise to a pair of underdamped compression/polarization waves propagating in opposite directions (as they have opposite imaginary parts) but decaying at the same rate  (see the inset of Fig.~\ref{FitDinamici}).
As $k$ increases, the real parts of $\lambda_0$ and $\lambda_\tau$ increase too.
The  mode associated with $\lambda_\gamma$ instead maintains its identity and stays real and positive, showing a slow decrease. 
Region III is again characterized by three distinct real modes:  $\lambda_0$ and $\lambda_\gamma$ monotonically decreasing, while $\lambda_\tau$ increases until it crosses $\lambda_\gamma$.
Finally, in region IV, the two modes associated with $\lambda_\gamma$ and $\lambda_\tau$ give birth to a pair of underdamped waves propagating in opposite directions and having a large adsorption rate, while the $\lambda_0$-mode is again real.

It is perhaps useful to recall the scenario in the passive case, $v_0=0$.
In this case the $\lambda_\tau$ mode is positive, completely decoupled, increases quadratically with $k$  and remains always real. The other two eigenvalues, say $\tilde \lambda_0$ and $\tilde \lambda_\gamma$ are real both below a threshold
$ k_1\approx \gamma/(2 v_s)$ and above the value $k_2=\frac{2 v_s}{\nu_\parallel}\sqrt{1-\nu_\parallel/2 v_s^2} $.
Between $k_1$ and $k_2$, the modes are complex and represent underdamped compression waves propagating in opposite directions. They result from the coupling between density and the momentum, like sound waves in ordinary liquids, 
and exist even in the presence of friction, $\gamma>0$. 
On the other hand, in active systems, the polarization-mode (the one corresponding to $\lambda_\tau=1/\tau$ at $k=0$) may couple with the other two modes and sustain propagating waves which have a much lower damping rate than the corresponding waves in the reference passive system.

\subsection{Transverse dynamical velocity-velocity structure function}

\begin{figure}[!t]
\centering
\includegraphics[width=1.\linewidth,keepaspectratio]{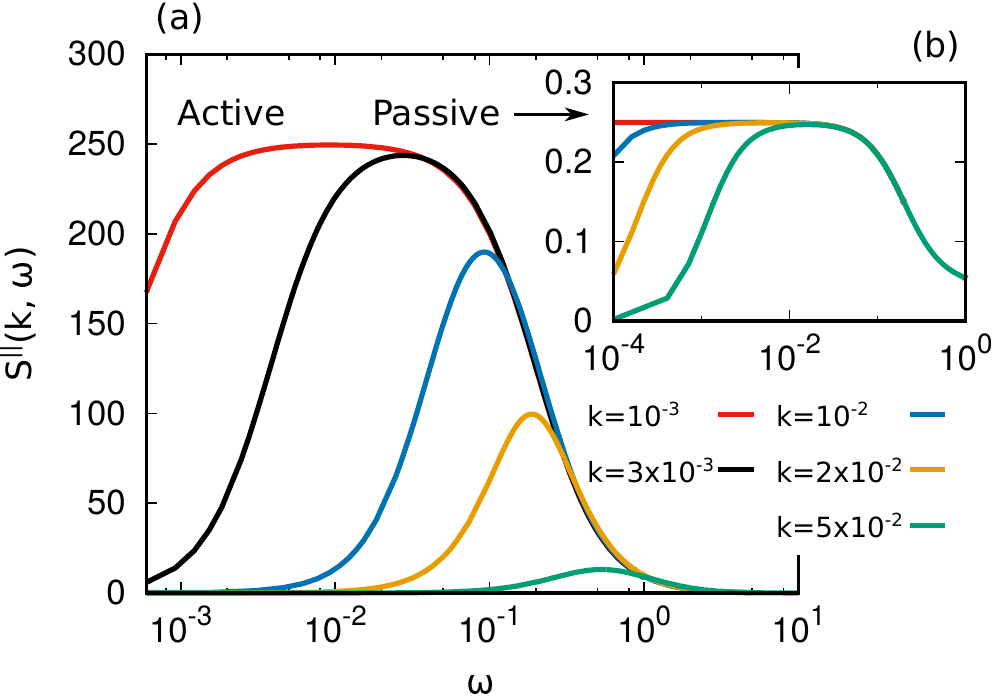}
\caption{\label{fig:SnapMag} 
Longitudinal velocity structure function $S^\parallel(k,\omega)$ versus $\omega$ for different values of $k$ for an active liquid (panel (a)) and a passive liquid (panel (b)). 
The values of the parameters are: $n_0=0.7$, $n_c=2$, $\tau=5$, $m=1$, $d=2$, $v_0=10$, $T_0=0.1$, $\gamma=4$,
with $v_0=10$ and $v_0=0$ in active and passive cases, respectively. 
}
\end{figure}

The transverse velocity-velocity dynamical structure factor, $S^\perp(k,\omega)$, can be easily obtained by solving Eq.~\eqref{dsf.5} with ${\bf M}^\perp$ (lower-right $2\times 2$ block of ${\bf M}$):
\begin{flalign}
 &S^\perp(k,\omega)= 2 v_0 v[n] \frac{\gamma^2 \lambda^\perp_\tau }{(\lambda^\perp_\gamma)^2-(\lambda^\perp_\tau)^2} \frac{1}{\omega^2+(\lambda^{\perp}_\tau)^2} \\
& + \left(\frac{2 T_0}{m}\lambda^{\perp}_\gamma  +2 v_0 v[n] \frac{\gamma^2 \lambda^\perp_\tau }{(\lambda^\perp_\tau)^2-(\lambda^\perp_\gamma)^2}\right)\frac{ 1}{\omega^2+(\lambda^{\perp}_\gamma )^2} \nonumber\,.
\label{e104}
\end{flalign}
This observable has a simple form being the sum of two Lorentzians each associated with one of the two transverse eigenvalues,
$\lambda^{\perp}_\gamma(k)$ and $\lambda^{\perp}_\tau(k)$ given by Eqs.~\eqref{eq:autotp} and~\eqref{eq:autotu}, respectively.
Multiplying by $e^{i\omega t}/2\pi$ and integrating with respect to $\omega$, we obtain the time-dependent intermediate scattering function:
 \begin{equation}
\begin{aligned}
F^\perp(k,t)=&\left[\frac{T_0}{m} -\frac{\gamma^2 v_0 v[n] }{(\lambda^\perp_\gamma)^2-(\lambda^\perp_\tau)^2}
 \frac{\lambda^\perp_\tau}{\lambda^\perp_\gamma} \right]e^{-\lambda^\perp_\gamma|t|}\\
&  +\frac{\gamma^2 v_0 v[n] }{(\lambda^\perp_\gamma)^2-(\lambda^\perp_\tau)^2} e^{-\lambda^\perp_\tau|t|}   \,,
\label{e104b}
\end{aligned}
\end{equation}
which decays as a linear combination of two time-exponentials, $e^{-\lambda^{\perp}_\gamma| t|}$ and $e^{-\lambda^{\perp}_\tau| t|}$, whose relative weight is mainly controlled by the ratio $\lambda^{\perp}_\tau / \lambda^{\perp}_\gamma$ (because $T_0$ is usually negligible).
The leading term in Eq.~\eqref{e104b} depends on the choice of the parameters: in the inertial active regime, $\tau\gamma\ll1$, the first term is dominant since $\lambda^{\perp}_\gamma \ll \lambda^{\perp}_\tau$ while, in  the overdamped active regime, $\tau \gamma \gg 1$, only the exponential with rate $\lambda^{\perp}_\tau$ survives because $\lambda^{\perp}_\gamma \gg \lambda^{\perp}_\tau$. Instead, in passive liquids,  where $v_0=0$, $F^\perp(k,t) \sim e^{-\lambda_\gamma t}$.
  This is shown in Fig.~\ref{figintermediate} where $F^\perp(k,t)$ is reported for different values of $k$ and the same choice of parameters as Fig.~\ref{FitDinamici} corresponding to the overdamped regime (in the inset of Fig.~\ref{figintermediate} the corresponding $S^\perp(k, \omega)$ is plotted as a function of $\omega$).
For active liquids ($v_0^2 \gg T_0/m$), our plot shows that the relaxation of $F^\perp(k,t)$ is mainly determined by $\lambda_\tau$, which is $\sim 1/\tau$ in the small-$k$ limit.
At variance with passive liquids, a $k$-dependence is observed in the active case in agreement with the correlation length of the spatial velocity correlation discussed in Sec.~\ref{equaltimecorrelations}.
We conclude that the intermediate transverse velocity structure function decays more slowly (at a rate $\sim 1/\tau$)
than the corresponding quantity in the passive case ($v_0=0$) whose decay rate is $\sim \gamma$, as illustrated in Fig.~\ref{figintermediate}.

\subsection{Longitudinal dynamical velocity-velocity structure function}
\label{struf}

  The discussion concerning the longitudinal modes is algebraically more involved than the one regarding the transverse modes, but 
follows similar lines.  By using Eq.~\eqref{dsf.5} with ${\bf M}^\parallel$ (the upper left $3\times 3$ block of $\mathbf{M}$ in \eqref{fh.1}), we can express the  density-density structure factor, $S_{nn}(k, \omega)$, in terms of the longitudinal eigenvalues $\lambda_0,\lambda_\tau,\lambda_\gamma$ according to the following formula:
\begin{widetext}
\begin{eqnarray}
 S_{nn}(k,\omega)=
2 n_0^2 k^2
\frac{ \Biggl[\frac{ T_0}{m}( \gamma+  \nu_{\parallel} k^2 )  
 \left( \omega^2+
 \left(\frac{1}{\tau}+ \frac{ T^*}{m \gamma} \frac{\tau\gamma}{1+\tau\gamma} k^2\right)^2 
 \right) 
+\gamma^2 v_0 v[n]\left(\frac{1}{\tau}+ \frac{ T^*}{m \gamma} \frac{\tau\gamma}{1+\tau\gamma} k^2\right)  
  \Biggr]}{ \Pi_{\alpha=1}^3 \left[\bigl(\omega+\Im(\lambda_\alpha)\bigr)^2+\bigl(\Re(\lambda_\alpha)\bigr)^2 \right]}
\label{snnkomega}
\end{eqnarray}
\end{widetext}
where $\Re(\lambda_\alpha)$ and $\Im(\lambda_\alpha)$ represent the real and imaginary part, respectively, of the
eigenvalue $\lambda_\alpha$ with $\alpha=(0, \gamma, \tau)$.
To gain better insight into the dynamics of the model, we inspect Eq.~\eqref{snnkomega}. In regions I and III of Fig.~\ref{FitDinamici},
$S_{nn}(k,\omega)$, as a function of $\omega$, is a linear combination of three Lorentzians.
Instead, in regions II and IV, the dynamic structure factor may in principle develop three distinct peaks at $\omega=0$ and $\omega\approx\pm \mathrm{Im} \left(\lambda_C\right)$ where  $\lambda_C$ is one of the two complex conjugate eigenvalues in regions II and IV.  
The width of these peaks is approximately given by $\left|\mathrm{Re}( \lambda_C)\right|$. However, when the damping is large ($\tau \gamma \gg 1$) well-separated peaks are hardly observable because the real part of the eigenvalues is very large. 

\begin{figure}[!t]
\centering
\includegraphics[width=0.9\linewidth,keepaspectratio]{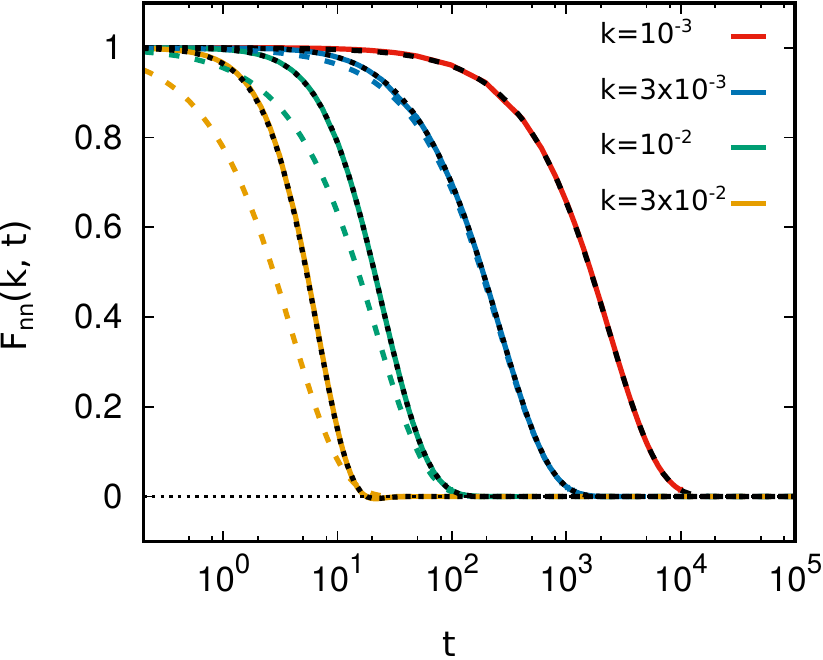}
\caption{\label{fig:Fnn} 
Longitudinal density intermediate scattering function $F_{nn}(k, t)$ (normalized with $F_{nn}(k,0)$) versus $t$ for different values of $k$ (solid lines). 
The dashed lines are obtained by plotting $e^{-\lambda_0 t}$ and colored according to the legend except for $k=10^{-3}$ where this curve is reported in black for presentation reasons. Dotted black lines (superimposed with the corresponding colored solid lines) are obtained by the prediction~\eqref{eq:F_nn} normalized to 1.
The values of the parameters are: $n_0=0.7$, $n_c=2$, $\tau=5$, $m=1$, $d=2$, $v_0=10$, $T_0=0.1$, $\gamma=4$ and $v_0=10$.
}
\end{figure}

Let us go back to the main object of our investigation, namely the longitudinal velocity-velocity correlation function $S^\parallel(k,\omega)$.
This observable can be obtained by Eq.~\eqref{snnkomega} through the general relation:
\begin{equation} 
S^\parallel(k,\omega)=\frac{1}{n_0^2}\frac{\omega^2}{k^2}S_{nn}(k,\omega)\,.
\end{equation} 
Since this formula depends on a variety of parameters, we shall limit ourselves to illustrate the behavior of  $S^\parallel(k,\omega)$ for a special selection as shown in Fig.~\ref{fig:SnapMag} (a) and (b) for active and passive particles, respectively.
We vary $\omega$ in correspondence of several values of $k$ and explore the regions represented in Fig.~\ref{FitDinamici}.
In the active case, after a small-$\omega$ regime which depends on $k$, a maximum in $\omega$ is approached and, then each $S^\parallel(k,\omega)$ shows a similar $k$-independent decay.
As $k$ increases, the height of the maximum as a function of $\omega$ decreases, and the peak slightly moves from $\sim 1/\tau$ towards larger values of $\omega$.
These small frequency peaks are not observed in the  passive case ($v_0=0$): there, for the same $k$-values, $S^\parallel(k,\omega)$ shows a rather flat maximum almost  $k$-independent.

\begin{figure}[!t]
\centering
\includegraphics[width=1.\linewidth,keepaspectratio]{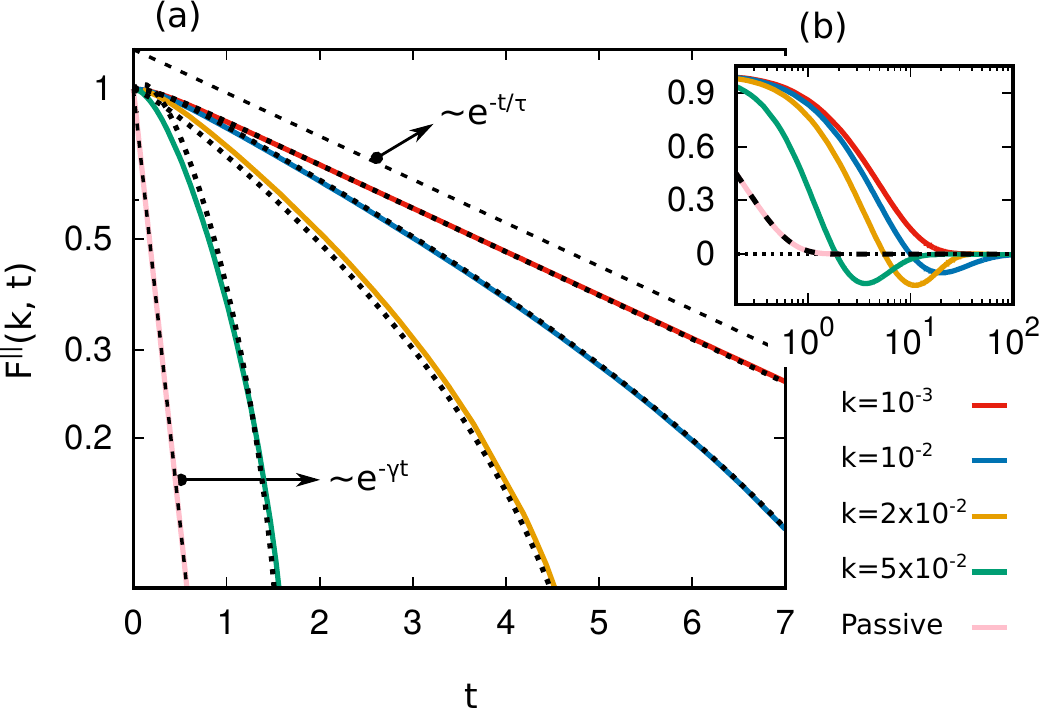}
\caption{\label{fig:Flong_long} 
 Longitudinal velocity intermediate scattering function $F^\parallel(k,t)$ (normalized with $F^\parallel(k,0)$) versus $t$ for different values of $k$ (colored lines). 
Panels (a) and (b) display the same observable in logarithmic and linear scale to outline different aspects of the time-decay. 
The two dashed black lines are guides for the eye showing the functions $\sim e^{-t/\tau}$ and $\sim e^{-\gamma t}$ as indicated in the graph.
Dotted black lines (superimposed with the corresponding colored solid lines) show the prediction~\eqref{eq:F_long}.
Finally, the pink solid line refers to the passive case and does not depend on $k$ for the value considered, $k=5\times 10^{-2}$.
The parameters are: $n_0=0.7$, $n_c=2$, $\tau=5$, $m=1$, $d=2$, $v_0=10$, $T_0=0.1$, $\gamma=4$ and $v_0=10$.
}
\end{figure}

The study of the intermediate scattering functions, $F_{nn}(k,t)$ and $F^\parallel(k,t)$, provides complementary information on the relaxation behavior of the fluctuations.
Let us begin by $F_{nn}(k,t)$: the location in the complex frequency plane of the poles in Eq.~\eqref{snnkomega} indicates that, in region I, this function can be expressed as a linear combination of three different temporal relaxations (each associated with one of the three Lorentzians mentioned above) characterized by exponential decays proportional to $e^{-\lambda_0(k) t}$, $e^{-\lambda_\gamma(k) t}$, $e^{-\lambda_\tau(k) t}$.
As illustrated in Appendix~\ref{app:intermediate}, the full calculation is performed by time-Fourier transforming the lengthy formula for  $S_{nn}(k,\omega)$.
Again, to proceed further, we restrict to the overdamped set-up analyzed so far. 
In Fig.~\ref{fig:Fnn}, we display $F_{nn}(k,t)$ versus $t$ for several values of $k$. We find that
for $\lambda_0 \ll \lambda_\tau \ll \lambda_\gamma$ and  small $k$, an approximate relation holds: 
\begin{equation}
\label{eq:F_nn}
 F_{nn}(k,t) \sim  e^{-\lambda_0 t} - \frac{\lambda_0}{\lambda_\tau} e^{-\lambda_\tau t} \,.
\end{equation}
The behavior~\eqref{eq:F_nn} is derived in Appendix~\ref{app:intermediate} and displayed in Fig.~\ref{fig:Fnn}.
The function $F_{nn}(k,t) $ is dominated by $e^{-\lambda_0(k) t}$ whose characteristic time diverges
diverges as $k\to 0$: as expected, the fluctuations of a conserved density relaxes diffusively towards the steady-state. 
The relaxation term associated with the polarization fluctuation provides the first correction to the expression for $F_{nn}(k,t)$ with relative weight $\sim \lambda_0/\lambda_\tau$ which becomes more relevant as $k$  increases (see Fig.~\ref{fig:Fnn}). 
On the contrary, such a contribution is absent in the passive case which is almost $k$-independent (not shown).

By time-Fourier transforming $S^\parallel(k,\omega)$, we compute the velocity-velocity intermediate scattering function, $F^\parallel(k,t)$, to shed light on the attenuation of the longitudinal velocity modes.
Again, the form of $S^\parallel(k,\omega)$ suggests that $F^\parallel(k,t)$ decays according to three exponential processes: $e^{-\lambda_\gamma(k) t}$, $e^{-\lambda_\tau(k) t}$ and $e^{-\lambda_0(k) t}$ but having different weights with respect to those featuring in $F_{nn}(k,t)$ (see Appendix~\ref{app:intermediate}.)
In Fig.~\ref{fig:Flong_long}, $F^\parallel(k,t)$ is shown as a function of $t$ for several values of $k$ both in the passive ($v_0=0$) and active overdamped cases (such that $\lambda_0 \ll \lambda_\tau \ll \lambda_\gamma$).
In these regimes, $F^\parallel(k,t)$ can be approximated as (see Appendix~\ref{app:intermediate}):
\begin{equation}
\label{eq:F_long}
 F^\parallel(k,t)\sim e^{-\lambda_\tau t}- \frac{\lambda_0}{\lambda_\tau} e^{-\lambda_0 t} - \frac{\lambda_\tau}{\lambda_\gamma} e^{-\lambda_\gamma t}  \,.
\end{equation}
As in the transverse case, $F^\parallel(k,t)$ shows an initial fast decay mainly due to the $e^{-\lambda_\gamma(k) t}$ contribution. 
At variance with passive suspensions, where this exponential is dominant and $F^\parallel(k,t)\approx F^\parallel(k,0) e^{-\lambda_\gamma t}$, in active suspensions this term has a small relative weight, $\propto \lambda_\tau/\lambda_\gamma$.
The leading contribution to  $F^\parallel(k,t)$ goes as $e^{-\lambda_\tau(k) t}$, but
at variance with $F^\perp(k, t)$,  one observes a stronger dependence on $k$: the larger the wavevector, the faster is the decay of $F^\parallel(k,t)$, an effect determined by the presence of the slowest exponential   $e^{-\lambda_0 t}$ of relative weight $\lambda_0/\lambda_\tau$ and negative sign. 
As its amplitude vanishes when $k\to 0$, its influence becomes completely negligible in the small-$k$ region but is
appreciable when $k\gtrsim 10^{-2}$ and is responsible for the negative values of $F^\parallel(k,t)$ for large times.

In conclusion, the analysis of the correlations using the $(k,t)$ space shows that the slowest modes associated with the eigenvalue $\lambda_\tau$ control the long-time behavior of the longitudinal velocity-velocity time correlation. 

\section{Conclusions}

In this paper, we have derived a simple hydrodynamic theory for interacting spherical active particles which applies to both ABP and AOUP underdamped systems, described by positions and velocities. 
We started from a microscopic level where the history of the system requires the knowledge of the full-space Fokker-Planck equation. 
Then, we coarse-grained the description switching to a hydrodynamic picture by taking the moments of the Fokker-Planck distribution function. While the application of this procedure is standard in many problems, in the case of the active models considered here, the choice of the appropriate hydrodynamic fields is not straightforward since the only conserved variable is the density and the assumption of strong scale-separation for the other modes (with respect to the fast degrees of freedom) cannot be rigorously guaranteed. 
Nevertheless, such a choice is crucial to observe the collective phenomena of interest.
At variance with the majority of previous approaches, we considered the balance equations not only
for the density and polarization but also for the momentum and temperature density fields. 
The interactions were accounted for by including the appropriate collisional terms which determine the so-called Irving-Kirkwood pressure and the transport coefficients of the model but produce also a decrease of motility of the particles, in agreement with previous works. 

In the second part of the paper, we investigated how the above equations can be used to predict the
fluctuations of the hydrodynamic fields about their average values by employing the methods of linear fluctuating Hydrodynamics.
We showed that, by taking into account the momentum field, one can observe, even in the case of large drag coefficients, spatially extended velocity correlations reminiscent of those experimentally and numerically observed in high-density active matter systems.
The interplay between persistent active driving and elastic response of the fluid to compression leads to correlations similar to those found in particle-based simulations and  theoretical investigations of solid-like configurations of active particles~\cite{caprini2020hidden, henkes2020dense, caprini2021spatial}.

The hydrodynamic theory predicts that the longitudinal velocity-velocity correlation function decays exponentially
at large distances with a characteristic length that is mainly determined by the sound speed, and is an increasing function of the persistence time of the active force (i.e. the inverse of the rotational diffusion coefficient), but a decreasing function of the solvent drag coefficient.
Surprisingly, we found that the time evolution of the velocity modes at a large scale is dominated by the coupling with polarization and not by the viscous damping, therefore their decay is substantially slower with respect to that of passive liquids. 
The shear transverse correlation function displays a correlation length that is almost independent of the active force persistence and mainly determined by the transverse viscosity: it is shorter than the longitudinal one because only depends on the shear rate but not on the compressibility of the liquid. 
Thus, our theory represents an extension to the liquid realm of the analogous treatment of velocity correlations in active solids.
Contrary to that case, the present study is not based on the equations of motion for the coordinates of the particles but considers the 
evolution of the collective hydrodynamic variables. 
This study shows that the velocity field is an important observable of active liquids, in contrast with
the widespread approach where one only considers an effective equation for the density field and thus disregards the velocity.
Despite the latter procedure has been successfully employed to understanding the onset of density inhomogeneity, it is legitimate to ask whether a wider picture as the one we presented  could shed some new light on the active phase separation.
Understanding the possible influence of the velocity ordering process on the structural properties of the system, such as the MIPS transition or the shift of liquid-hexatic and hexatic solid transitions of homogeneous phases~\cite{bialke2012crystallization, digregorio2018full, mandal2020extreme, caprini2020hidden}, still represents an open question. 

\begin{acknowledgments}
The authors acknowledge financial support from MIUR through the PRIN 2017 grant number 201798CZLJ. AP also acknowledges financial support from Regione Lazio through the Grant “Progetti Gruppi di Ricerca” N. 85-2017-15257.
\end{acknowledgments} 

  \appendix

\section{Parameter estimate} \label{app:Parameter}

To evaluate the system's properties it is necessary to specify the equation of state and the transport coefficients.
We begin with the static contribution to the pressure tensor, $P_h$.
Such a quantity accounts for the so-called ideal and repulsive contributions to the pressure.
A reasonable form is the hard-disks equation of state 
of Henderson~\cite{henderson1975simple}:
\begin{equation}
P_h=n  T^*\left[ \frac{1+y^2/8}{(1-y)^2}\right] \,,
\end{equation}
where $y=\frac{\pi}{4} n\sigma^2 $ represents the packing fraction.
The isothermal sound speed is  simply obtained from the pressure through the relation:
\begin{equation}
v_s^2=\frac{B}{mn}= \frac{1}{m}\frac{\partial P}{\partial n} \,,
\end{equation}
where $B$ is the bulk modulus. Explicitly we find:
\begin{equation}
v_s^2=\frac{ T^*}{m} 
\Bigl[\frac{1+y+3  y^2/8- y^3/8}{(1-y)^3} \Bigr] \,.
\end{equation}We also need the expressions for the  dynamic shear viscosity:
\begin{equation}
\eta=\eta_0\left[ \frac{1}{g_2(y)}+2y+\left(1+\frac{8}{\pi}\right) y^2g_2(y)\right]
\end{equation}
and the bulk viscosity~\cite{gass1971enskog}
\begin{equation}
\zeta= \frac{16}{\pi} y^2 g_2(y) \eta_0 \,,
\end{equation}
with
\begin{equation}
\eta_0=\frac{1}{2 \sigma} \sqrt{ \frac{m  T^*}{\pi}} \,.
\end{equation}
The formulas contain the pair correlation function at contact distance between two disks, $g_2(y)$, which in the Verlet-Levesque approximation can be estimated as:
\begin{equation*}
g_2(y)=\frac{1-7y/16}{(1-y)^2} \,.
\end{equation*}
Finally, the kinematic viscosity is obtained by dividing $\eta$ by $m n_0$ (assuming that the density is constant)
\begin{equation}
\nu_0=\frac{\eta_0}{m n_0}=\frac{1}{2 \sigma n_0} \sqrt{  \frac{T^*}{m \pi }}  \,.
\end{equation}
Notice that $\nu_0$ diverges as density $n_0\to 0$, because in the absence of collisions there is no diffusion of momentum, but the phenomenon becomes ballistic. 
Now, the expressions for $\nu_\perp$ and $\nu_\parallel$ can be obtained as:
\begin{flalign}
&
\nu_{\perp}=\frac{\eta}{m n_0} \\
&
\nu_{\parallel}=\frac{1}{mn_0 }\left( \frac{4}{3}\eta+\zeta\right)\label{nuparallel} \,.
\end{flalign}
These explicit expressions of the transport coefficients 
have been employed in the numerical study.

\section{Effective decoupling of thermal and density fluctuations}
\label{app:demonstration}

 In this appendix, we justify the assumption of neglecting the influence of the temperature fluctuation
on the remaining fields and discarding Eq.\eqref{eq:lin_temperature}. This is possible because the coupling between velocity and temperature fields is proportional to the thermal expansion coefficient 
so that  its effect is rather small on the evolution of the other variables.

 The first two terms in the r.h.s. of Eq.~\eqref{eq:lin_velocitylongitudinal} can be written as
\begin{equation}
\begin{aligned}
  &\frac{(\partial P/\partial n)_T}{m n_0 }   \mathbf{\nabla} n+ \frac{(\partial P/\partial T)_n}{m n_0 } \mathbf{\nabla} T \\
  &\qquad\qquad\qquad=  \frac{v_s^2}{n_0}   \mathbf{\nabla} n + \alpha v_s^2 \mathbf{\nabla} T  \,,
\label{rolethermal}
\end{aligned}
\end{equation}
where the thermal expansion coefficient, $\alpha$, is defined as:
 \begin{equation}
\alpha=-\frac{1}{n} \left[ \frac{\partial n}{\partial T} \right]_P \,.
\end{equation}
In Eq.~\eqref{rolethermal}, the first term dominates over the second since the liquid pressure increases more slowly with respect to temperature than with respect to density changes, at least for the high-packing regime considered in this work. 
This can be shown by considering the ratio between the two amplitudes in Eq.~\eqref{rolethermal} as a function of the packing fraction:
\begin{equation}
\frac{(\partial P/\partial T)_n}{(\partial P/\partial n)_T}=\frac{n(1+y^2/8)(1-y)} {T^*(1+y+3  y^2/8- y^3/8)} \,.
\end{equation}By the same token, when
  $y \to 1$ the transport coefficients $\nu_\parallel$ and $\kappa$ are less divergent than the term $v_s^2$ , as evident from \eqref{nuparallel}.
 This means that $v^2_s$ is the dominant transport coefficient in the expression for $\xi_\parallel$, Eq.~\eqref{eq:xi_parallel}.


\section{Noise amplitudes}\label{app:noise_amplitude}

To introduce our method to determine the noise amplitudes, one can think that the overall noise acting on each mode results
from the combined effect of the external noise (the solvent) and of the fast (non-hydrodynamic modes) which have been eliminated by our coarse-graining procedure, but still exert their influence.
 A formal and elegant justification of the assumed formula for ${\bf D}$ requires some discussion of what is
external and internal noise, a problem studied in  Refs.~\cite{van1999randomly,gradenigo2011fluctuating}.
In practice, we will fix the noise amplitudes ${\bf D}$ by requiring the following two conditions: 1) the velocity correlations must reduce to their known equilibrium form in the passive limit,  $v_0\to0$ and 2) when the system is uniform and $v_0> 0$, the expression for ${\bf D}$ should be consistent with the expressions found in Sec.~\ref{eq:app_uniform}.

\subsection{ Equal-time transverse velocity correlations}

The expression of the transverse equal-time velocity correlation can be obtained solving the $2\times2$ problem, Eq.~\eqref{structurefactors}, in the Fourier space:
\begin{equation}  
\delta\dot{\bf a}_\perp({\bf k},t)=-{\bf M}^\perp(k)\delta{\bf a}_\perp({\bf k},t),
\end{equation}
where $\hat{\bf a}_\perp = \{\hat{u}_\perp({\bf k},t), \hat{p}_\perp({\bf k},t)\}$ and the matrix ${\bf M}^\perp(k)$ reads:
\begin{equation}
\mathbf{M}^\perp = 
\left( \begin{array}{cccccc}
\gamma+  \nu_{\perp} k^2 &-\frac{1}{n_0}\gamma v_0 \\ 
0&\frac{1}{\tau}+ \frac{ T^*}{m \gamma} \frac{\tau\gamma}{1+\tau\gamma} k^2 
\end{array} \right) \,.\nonumber\\
\label{fh.1tris}
\end{equation}
%
%
 Straightforward calculations lead to the following form of the equal-time correlations:
\begin{eqnarray}&&
\label{cperpuu}
C^{\perp}_{uu}= \frac{D^{\perp}_{uu}}{M^{\perp}_{uu}} + \frac{D^{\perp}_{pp} (M^{\perp})^2_{up}}{M^{\perp}_{uu}M^{\perp}_{pp}(M^{\perp}_{uu}+M^{\perp}_{pp} )} \\&&
C^{\perp}_{pp}=\frac{D^{\perp}_{pp}}{M^{\perp}_{pp}} \\&&
C^{\perp}_{up}=-\frac{D^{\perp}_{pp}}{M^{\perp}_{pp}}  \frac{M^{\perp}_{up}}{(M^{\perp}_{uu}+
M^{\perp}_{uu}  )}  \,.
\end{eqnarray}
Now, to fix $D^{\perp}_{uu}$  
we require that  in the limit $v_0\to 0$ the velocity correlations are equal to those of the reference passive system by imposing the following condition:
\begin{equation}\frac{D^{\perp}_{uu}}{M^{\perp}_{uu}}=  \frac{T_0}{m}\, .
\end{equation}
In addition, to obtain the correct matching when $k=0$ between the velocity fluctuation in the uniform active system given by Eq.~\eqref{eq:corr_matrx_vv_interact} and Eq.~\eqref{cperpuu}, we assume:
\begin{equation}
\frac{D^{\perp}_{pp}}{M^{\perp}_{pp}}= n_0^2 \frac{v[n]}{v_0}
\end{equation}%
and by substituting, we find
\begin{equation*}
C^\perp= \frac{T_0}{m} + \frac{ \gamma^2 v_0 v[n]}{\left[\gamma+\nu_\perp k^2  \right] \left(\left[\gamma+\nu_\perp k^2  \right]+
\left[\frac{1}{\tau}+\frac{\tau\gamma}{1+\tau\gamma} \frac{ T^*}{m \gamma} k^2 \right]\right)} \, .
\end{equation*}
Keeping only $k^2$ orders in the denominator, we obtain:
 \begin{equation}
C^\perp \approx \frac{T_0}{m} + \frac{  v_0 v[n]}{1+\frac{1}{\tau \gamma}}\frac{1}{1+\xi^2_\perp k^2} \,,
\label{app:transversecorrelation}
\end{equation}
where $\xi_\perp$ is the correlation length associated with the transverse mode and is given by Eq.~\eqref{eq:xi_perp}.

\subsection{Equal-time longitudinal velocity correlations}
We, now, apply the same reasoning to the longitudinal correlations.
Incidentally, we recall that $C_{\psi\psi} $ and  $C_{\phi\phi} $ are the correlations
 of $\psi=ik u_k$ and $\phi=i k p_k$, respectively (see Eq.~\eqref{campipsiphi}).
 The passive form of the correlations is
\begin{eqnarray}&&
C^{passive}_{\psi\psi} =\frac{D_{\psi\psi}}{M_{\psi \psi} }= \frac{ T_0}{m} k^2\\&&
 C^{passive}_{\psi n}=0\\&&
 C^{passive}_{nn}=-\frac{M_{n\psi}}{M_{\psi n}} \frac{D_{\psi\psi}}{M_{\psi\psi} } = T_0 \frac{n_0^2}{ v_s^2} \, ,
 \end{eqnarray}
so that we deduce the following amplitude:
 \begin{equation}
D_{\psi\psi}= M_{\psi\psi}\frac{ T_0}{m} k^2 =\frac{ T_0}{m} k^2( \gamma+  \nu_{\parallel} k^2 )
\label{eqdpsipsi}
 \end{equation} 
 Regarding the amplitude associated with the polarization noise we solved Eq.~\eqref{structurefactors} for $C^\parallel(k)$
 and matched its solution with the homogeneous result ($k=0$) given by Eq.~\eqref{eq:corr_matrx_vv_interact}.

We consider the reduced $3\times3$ problem (for $\hat{\bf a}_\parallel=\{\delta \hat{n}({\bf k},t), \hat{\psi}({\bf  k},t),  \hat{\phi}({\bf k},t)\}$)
\begin{equation}  
\frac{d}{dt}\delta\hat{\bf a}_\parallel({\bf k},t)=-{\bf M}^\parallel(k)\delta \hat{{\bf a}_\parallel}({\bf k},t),
\end{equation}
where the matrix ${\bf M}^\parallel(k)$ reads:
\begin{widetext}
\begin{equation}
\hspace{-2.5cm}\mathbf{M}^\parallel = 
\left( \begin{array}{cccccc}
0 & n_0 & 0\\ 
 - \frac{v_s^2}{n_0} k^2  &\gamma+  \nu_{\parallel} k^2  &  -\frac{1}{n_0}\gamma v_0  \\ 
- \frac{\tau\gamma}{1+\tau\gamma} (v[n]-v_0  \frac{n_0}{n_c}) k^2  &  \frac{2}{\gamma}(\frac{\tau\gamma}{1+\tau\gamma} )^2 n_0 v[n]   k^2 &\frac{1}{\tau}+ \frac{ T^*}{m \gamma} \frac{\tau\gamma}{1+\tau\gamma} k^2&\\ 
\end{array} \right) \,.\nonumber\\
\label{fh.1bis}
\end{equation}
 After tedious algebra we obtain an approximate expression that captures the leading contributions:
 \begin{equation}C_{\psi\psi}(k) \approx\frac{ D_{\psi\psi}}{ M_{\psi\psi} } +   \frac{1}{M_{\psi\psi}}
 \frac{M_{\psi\phi}^2 D_{\phi\phi}  }{ \Bigl((M_{\psi\psi} +M_{\phi\phi})M_{\phi\phi}  -  M_{\psi\phi}M_{\phi\psi}  -M_{\psi n} M_{n\psi}    \Bigr)} \,.
\label{implicitab1}
\end{equation}\end{widetext}
Now, to match $C^\parallel(k)= \frac{C_{\psi\psi}(k)}{k^2}$  with the expression Eq.~\eqref{eq:corr_matrx_vv_interact} when $k=0$ is sufficient to choose
 \begin{equation}
 D_{\phi\phi}= n_0^2 M_{\phi\phi} k^2=n_0^2\Bigl(\frac{1}{\tau}+ \frac{ T^*}{m \gamma} \frac{\tau\gamma}{1+\tau\gamma} k^2\Bigr) \, \frac{v[n]}{v_0}k^2
 \end{equation}
Replacing the expression for the elements of ${\bf M}^\parallel$, the noise coefficients, and keeping only orders $k^2$ in the denominator of Eq.~\eqref{implicitab1}, we  obtain:
\begin{equation}C^\parallel(k)=\frac{T_0}{m}  +\frac{ v^2_0 }{1+\xi^2_\parallel k^2} \,,
 \label{app:velocityfluctuation}
  \end{equation}
where $\xi_\parallel$ is the correlation length associated to the longitudinal modes that is given by  formula~\eqref{eq:xi_parallel}.

\section{Calculation of the intermediate scattering functions}
\label{app:intermediate}
In order to calculate the intermediate scattering functions,
we consider the following Fourier transforms obtained by Cauchy's integral formula:
\begin{equation}R_{2n}(t)=
 \int_{-\infty}^\infty \frac{d\omega }{2\pi}  \,\frac{\omega^{2n}\, e^{i\omega t} }{(\omega^2+\lambda_0^2)(\omega^2+\lambda_\gamma^2)(\omega^2+\lambda_\tau^2)} \nonumber \,.
\end{equation}
\begin{widetext}
For real eigenvalues, the functions $R_{2n}(t)$ for $n=0,1,2$ esplicitly reads:
\begin{eqnarray}
&&
R_0(t)=\frac{1}{2}\frac{1}{\lambda_\gamma^2-\lambda_0^2} \Biggl[  \frac{1}{\lambda_\tau^2-\lambda_0^2} \Bigl(\frac{1}{ \lambda_0 } e^{-\lambda_0 |t|}-\frac{1 }{\lambda_\tau} e^{-\lambda_\tau|t|}  \Bigr) 
  -  \frac{1}{\lambda_\tau^2-\lambda_\gamma^2} \Bigl(\frac{1 }{\lambda_\gamma } e^{-\lambda_\gamma |t|}-\frac{ 1 }{\lambda_\tau} e^{-\lambda_\tau |t|}  \Bigr) \Biggr]   \nonumber\\
&&
 R_2(t)=-\frac{1}{2}\frac{1}{\lambda_\gamma^2-\lambda_0^2} \Biggl[  \frac{1}{\lambda_\tau^2-\lambda_0^2} \Bigl(\lambda_0 e^{-\lambda_0 |t|}-\lambda_\tau e^{-\lambda_\tau|t|}  \Bigr) 
  -  \frac{1}{\lambda_\tau^2-\lambda_\gamma^2} \Bigl(\lambda_\gamma  e^{-\lambda_\gamma |t|}- \lambda_\tau  e^{-\lambda_\tau |t|}  \Bigr) \Biggr]   \nonumber\\
&&
R_4(t)=\frac{1}{2}\frac{1}{\lambda_\gamma^2-\lambda_0^2} \Biggl[  \frac{1}{\lambda_\tau^2-\lambda_0^2} \Bigl(\lambda_0^3 e^{-\lambda_0 |t|}- \lambda_\tau^3  e^{-\lambda_\tau|t|}  \Bigr) 
  -  \frac{1}{\lambda_\tau^2-\lambda_\gamma^2} \Bigl(\lambda_\gamma^3 e^{-\lambda_\gamma |t|}-\lambda_\tau^3 e^{-\lambda_\tau |t|}  \Bigr) \Biggr]   \nonumber
\end{eqnarray}
In region II and in region IV
two eigenvalues are complex conjugate ($\lambda_R\pm i\lambda_I$) and the third eigenvalue, $\lambda_3$ is real and positive, the Fourier transform for $t>0$ is:
\begin{eqnarray}
R_0(t)=&& \int_{-\infty}^\infty \frac{d\omega}{2\pi}   \frac{e^{i\omega t}}{\Bigl[\bigl(\omega+\lambda_I\bigr)^2+\lambda_R^2 \Bigr] \Bigl[\bigl(\omega-\lambda_I\bigr)^2+\lambda_R^2 \Bigr]
\Bigl[\omega^2+\lambda_3^2 \Bigr]}
=\frac{1}{\Bigl[(\lambda_I^2+\lambda_R^2-\lambda_3^2)^2+ 4 \lambda_I^2 \lambda_3^2 \Bigr] 
2\lambda_3}\, e^{-\lambda_3 t} \nonumber
\nonumber\\
&&
+\frac{1}{4\lambda_I \lambda_R(\lambda_I^2+\lambda_R^2)}\frac{\lambda_I \Bigl[\lambda_I^2+\lambda_3^2-3\lambda_R^2\Bigr] \cos(\lambda_I t)+\lambda_R \Bigl[3\lambda_I^2+\lambda_3^2-\lambda_R^2\Bigr] \sin(\lambda_I t)}
{(\lambda_I^2+\lambda_3^2-\lambda_R^2)^2 +4 \lambda_I^2 \lambda_R^2}e^{-\lambda_R t}\nonumber
\end{eqnarray}
and we obtain $R_{2n}(t)$ for $n>0$ by using the relation $R_{2n}(t)=(-1)^n \frac{d^{2n} R_0(t)}{dt^{2n}}$.
By Fourier transforming $S_{nn}(k,\omega)$ and $S^\parallel(k,\omega)$, 
we obtain the following expressions for the intermediate scattering functions, $F_{nn}( k,t)$ and $F^\parallel(k,t)$:
\begin{eqnarray}&&
 F_{nn}(k,t)
= 2 n_0^2 k^2 \Biggl\{  \frac{ T_0}{m} \left( \gamma+  \nu_{\parallel} k^2 \right)  R_2(t) 
     +\, \left(\frac{1}{\tau}+ \frac{ T^*}{m \gamma} \frac{\tau\gamma}{1+\tau\gamma} k^2\right)\left[ \frac{ T_0}{m} ( \gamma+  \nu_{\parallel} k^2 )
 \left(\frac{1}{\tau}+ \frac{ T^*}{m \gamma} \frac{\tau\gamma}{1+\tau\gamma} k^2\right)
+\gamma^2 v_0^2  \right] R_0(t)
  \Biggr\} \nonumber\\&&
\label{Fnnkomegabis}
\end{eqnarray}
and
\begin{equation}
\hspace{-0.1cm} F^\parallel(k,t)
=  2\Biggl\{  \frac{ T_0}{m} \left( \gamma+  \nu_{\parallel} k^2 \right)  R_4(t) 
  +\, \left(\frac{1}{\tau}+ \frac{ T^*}{m \gamma} \frac{\tau\gamma}{1+\tau\gamma} k^2\right)\left[ \frac{ T_0}{m} ( \gamma+  \nu_{\parallel} k^2 )
 \left(\frac{1}{\tau}+ \frac{ T^*}{m \gamma} \frac{\tau\gamma}{1+\tau\gamma} k^2\right)
+\gamma^2 v_0^2  \right] R_2(t)
  \Biggr\} 
\label{Fuukomegabis}
\end{equation}

\end{widetext}

\subsection{Scattering functions in the overdamped regime}

It is useful to extract the dominant terms in the  expressions for $F_{nn}(k,t)$ and $F^\parallel(k,t)$ in the regime of parameters reported in the figures to get approximated but explicit expressions.
Assuming the condition $v_0^2 \gg T_0/m$, we can neglect the term containing $R_2(t)$ in the expression for $F_{nn}(k,t)$ and the one containing $R_4(t)$ in the expression for $F^\parallel(k,t)$.
In addition, in the overdamped regime $\tau \gamma \gg 1$, we have $\lambda_0 \ll \lambda_\tau \ll \lambda_\gamma$ so that we can simplify the expressions for $R_{0}(t)$ and $R_2(t)$ as follows:
\begin{flalign}
\label{eq:app_R0_approx}
&R_{0}(t) \approx K_0 \left[ e^{-\lambda_0 |t|}   -\frac{\lambda_0}{\lambda_\tau} e^{-\lambda_\tau |t|}  +\frac{\lambda_0\lambda_\tau^2}{\lambda_\gamma^3}  e^{-\lambda_\gamma |t|}  \right] \\
\label{eq:app_R2_approx}
&R_{2}(t) \approx K_2 \left[ e^{-\lambda_\tau |t|} -\frac{\lambda_0}{\lambda_\tau} e^{-\lambda_0 |t|}  - \frac{\lambda_\tau}{\lambda_\gamma}  e^{-\lambda_\gamma |t|} \right]
\end{flalign}
where $K_0=K_0(k)$ and $K_2=K_2(k)$ are two k-dependent amplitudes. In Eqs.~\eqref{eq:app_R0_approx} and~\eqref{eq:app_R2_approx}, after expanding the relative weights of the different terms in powers of $\lambda_\tau/\lambda_\gamma \ll 1$ and $\lambda_0/\lambda_\tau \ll 1$, we have kept only the leading contributions.
Using these two expressions, we arrive at the relations leading to Eqs.~\eqref{eq:F_nn} and~\eqref{eq:F_long}
\begin{flalign}
& F_{nn}(k,t)\approx \mathcal{N}_n \left[ e^{-\lambda_0 |t|}   -\frac{\lambda_0}{\lambda_\tau} e^{-\lambda_\tau |t|}  +\frac{\lambda_0\lambda_\tau^2}{\lambda_\gamma^3}  e^{-\lambda_\gamma |t|}  \right] \nonumber \\
&F^\parallel(k,t)\approx \mathcal{N}_\parallel \left[ e^{-\lambda_\tau |t|} -\frac{\lambda_0}{\lambda_\tau} e^{-\lambda_0 |t|}  - \frac{\lambda_\tau}{\lambda_\gamma}  e^{-\lambda_\gamma |t|} \right]\nonumber
\end{flalign}
where $\mathcal{N}_n$ and $\mathcal{N}_\parallel$ are constants depending on $k$ through $K_0$ and $K_2$ and contain the prefactors present in formulas~\eqref{Fnnkomegabis} and~\eqref{Fuukomegabis}.
It is remarkable that the leading term in the expression for $F^\parallel(k,t)$ qualitatively agrees with the prediction obtained from a one-dimensional active solid~\cite{caprini2020time}.

\bibliographystyle{apsrev4-1}

\bibliography{correlazioniidrodinamiche.bib}

\end{document}